\documentclass[
  aps,
  prd,
  11pt,
  superscriptaddress,
  notitlepage,
  onecolumn,
  nofootinbib
]{revtex4-1}

\AtBeginDocument{
  \linespread{1.05}\selectfont
}

\usepackage[
  a4paper,
  left=2.3cm,
  right=2.3cm,
  top=2.5cm,
  bottom=2.5cm
]{geometry}

\usepackage{microtype} 
\usepackage{graphicx}
\usepackage[svgnames,dvipsnames]{xcolor}
\usepackage{orcidlink}
\usepackage{enumerate}
\usepackage{enumitem}
\usepackage{ulem}
\usepackage{amssymb} 
\usepackage{tikz}
\usepackage{mathtools}
\usepackage{ytableau}
\usepackage{soul}
\usetikzlibrary{arrows.meta,positioning,fit,calc}

\definecolor{linkcolor}{rgb}{0.0,0.3,0.5}
\definecolor{mypurple}{RGB}{143, 116, 210}

\graphicspath{{Figs/}}

\newcommand{\hias}{School of Fundamental Physics and Mathematical Sciences, Hangzhou Institute for Advanced Study, University of Chinese Academy of Sciences, Hangzhou 310024, China}
 
\begin{document}

\title{From Modal Memory to Topological Relaxation in a Driven Holographic Superfluid}

\author{Yan Peng
\orcidlink{0009-0004-7100-1307}}
\affiliation{\hias}
\author{Bo-Xuan Ge
\orcidlink{0000-0003-0738-3473}}
\email{bo-xuan.ge@ucas.ac.cn}
\affiliation{\hias}

\begin{abstract}
Different driving histories can leave a holographic superfluid
with different relative amplitudes in an unstable-mode pair,
even when the subsequent drive is identical. In an eleven-run
scan, all trajectories undergo five successive unit winding
changes and reach the same observed final sector,
$N_f=-5$, yet the residence time in the $N=-4$ sector varies
by a factor of approximately $3.6$. Within this history
family, the variation is well described by an empirical
exponential dependence on the arrival imbalance of an
unstable-mode pair with nearly equal growth rates. Distinct
protocols with matched imbalance give nearly equal lifetimes.
The imbalance is not a complete descriptor, however:
rotating the initial phase of the seeded $m=-7$ mode produces
a $19.2\%$ difference in the $N=-4$ lifetime while leaving
the primary-pair arrival imbalance unchanged. Thus, phase
information in the prepared multimode state also shapes
topological relaxation.
\end{abstract}

\maketitle
\section{Introduction}
\label{sec:introduction}

A phase slip changes the winding of a superflow, but it need not
remove the imprint of how the flow was prepared. Different drives
can bring a system to the same instantaneous boundary sources while
leaving different complex amplitudes in its unstable modes. The
question we address is whether this difference remains visible in
the nonlinear relaxation that follows.

Above the Landau critical velocity, counter-propagating excitations
can become dynamically unstable \cite{Landau1941,Amado2014}.
Holographic superfluids provide a useful setting in which to study
this process at strong coupling \cite{Herzog2009,Arean2010}. The
relation between dynamical and thermodynamic instabilities has been
examined in relativistic superfluid hydrodynamics
\cite{Gouteraux2023,Arean2024}, while nonlinear simulations have
revealed solitonic structures during the decay of unstable superflows
\cite{Keranen2010,Yang2021,Lan2025}. In a ring, phase slips connect
distinct winding sectors and provide a route for the relaxation of
persistent flow \cite{Varoquaux2015,LiShiZhang2022}. Recent
holographic simulations have also linked the structure of unstable
modes to the winding content of the resulting vortex configurations
\cite{Yan2026}. The time spent in an intermediate sector is therefore
a natural measure of the persistence of a metastable superflow.

Quasinormal modes (QNMs) provide the linear collective spectrum of
the symmetry-broken phase \cite{Amado2009}. In our previous work
\cite{Ge:2026nyg}, we showed that finite-rate driving can leave
different, and even opposite, relative amplitudes in competing
Landau-unstable modes despite identical instantaneous boundary
driving conditions. For the protocols studied there, this modal
memory was well described by the accumulated difference in their
QNM growth rates. That analysis concerned the amplification of weak
perturbations, before the evolution became strongly nonlinear. It
left open whether the prepared modal state could influence the
subsequent sequence of topological phase slips.

Here we follow different driving histories into the same subsequent
protocol. In the eleven-run primary history scan, all trajectories
undergo five unit changes of the winding number $N$ and reach the
same observed final winding, $N_f=-5$. The first slip occurs at
almost the same time, but the residence time in the penultimate
sector, $N=-4$, varies substantially. At higher spatial resolution,
the two endpoint histories give residence times of $26.00$ and
$95.10$, a factor of about $3.66$. Within this history family, the
variation is well described by an empirical exponential in the
arrival imbalance of the nearly degenerate primary mode pair.
A single dwell, a split dwell and a continuous slow window,
constructed to arrive with nearly the same imbalance, give nearly
identical residence times.

The amplitude ratio is not, however, a complete description of the
prepared state. Redistributing the initial amplitudes of the primary
pair reverses the trend found in the history scan. More strikingly,
rotating the initial phase of the seeded $m=-7$ mode produces a
$19.2\%$ difference in the $N=-4$ residence time while leaving the
primary-pair arrival imbalance unchanged to numerical precision.
The modal imbalance therefore organises the original history family,
but the subsequent kinetics also depends on phase information
carried by the other modes.

We examine one route for this phase information to enter the
nonlinear evolution. A Fourier decomposition identifies a
phase-sensitive contribution involving three neighbouring modes
in the condensate-density harmonic, whose bulk counterparts
enter the Maxwell source. This is an allowed coupling channel, not
evidence that one triad or source term dominates the dynamics.
We also vary the common-future velocity, test a regime in which
the leading competing pair changes, and repeat the phase
intervention on a second charge-density background. The response
persists in the cases examined, although its effect on the residence
time depends on which winding sector is measured. An independent
dissipative Gross--Pitaevskii model provides a further example:
two initial states that differ only in the phase coordinate
$\Phi_{567}$, but have matched global quantities, can have
different winding-sector lifetimes.

The paper is organised as follows. Section~\ref{sec:setup}
introduces the holographic model and boundary observables.
Section~\ref{sec:protocols} describes the QNM spectrum, mode
selection and preparation protocols. Section~\ref{sec:history_relaxation}
presents the history-dependent topological relaxation, while
Sec.~\ref{sec:controlled_interventions} examines the limits of
the amplitude-imbalance description through controlled modal
interventions. Section~\ref{sec:local_phase_slip} studies the
local phase-slip structure and Maxwell--scalar feedback.
Section~\ref{sec:robustness_generality} examines changes of the
common future, competing modes and background. Section~\ref{sec:conclusion}
concludes with a discussion of the results. Numerical protocols
and convergence tests are given in Appendix~\ref{app:convergence};
additional topological diagnostics and nonlinear checks in
Appendix~\ref{app:nonlinear_checks}; the Maxwell-source
decomposition in Appendix~\ref{app:maxwell_feedback}; and the
independent dissipative Gross--Pitaevskii calculation in
Appendix~\ref{app:dgpe_bridge}.

\section{Holographic Model and Observables}
\label{sec:setup}

\subsection{The bulk model}

We consider a holographic superfluid described by the
probe-limit Abelian--Higgs model on a planar
Schwarzschild--AdS$_4$ background
~\cite{Herzog2009,Arean2010,Ge:2026nyg}. The matter
action is
\begin{equation}
S_{\mathrm m}
=
\int d^4x\,\sqrt{-g}\,
\left[
-\frac{1}{4}F_{\alpha\beta}F^{\alpha\beta}
-(D_\alpha\Psi)^*D^\alpha\Psi
-m_\Psi^2|\Psi|^2
\right],
\label{eq:matter_action}
\end{equation}
where \(A_\alpha\) is the gauge field, \(\Psi\) is a
complex scalar, and \(g\) is the determinant of the
spacetime metric. We define
\(F_{\alpha\beta}=\partial_\alpha A_\beta-\partial_\beta A_\alpha\)
and \(D_\alpha=\nabla_\alpha-iA_\alpha\), with \(i\)
the imaginary unit and Greek indices denoting bulk
spacetime coordinates. The scalar charge, AdS radius
and horizon radius are set to unity, and the scalar
mass is \(m_\Psi^2=-2\). The matter fields do not
backreact on the geometry.

We use ingoing Eddington--Finkelstein coordinates,
in which the background metric reads
\begin{equation}
ds^2
=
\frac{1}{z^2}
\left[
-f(z)dt^2-2\,dt\,dz+dx^2+dy^2
\right],
\qquad
f(z)=1-z^3 .
\label{eq:ef_metric}
\end{equation}
The AdS boundary and the future horizon lie at
\(z=0\) and \(z=1\), respectively. The Hawking
temperature is \(T_{\mathrm H}=3/(4\pi)\). We restrict
to fields independent of \(y\), with periodic boundary
conditions along \(x\) over a length \(L_x=16\pi\). We denote
the number of spatial Fourier points by \(N_x\) and the number
of radial Chebyshev--Lobatto collocation points by \(N_z\).

We work in the canonical ensemble at fixed spatially
averaged charge density
\(\bar\rho=10.0417\), corresponding to
\(T/T_c\simeq0.637\), where \(T_c\) is the critical
temperature at the same charge density.

\subsection{Boundary observables and superfluid flow}

Near the AdS boundary, the fields have the expansions
\begin{equation}
\Psi=z\Psi_-+z^2\Psi_++\cdots,
\qquad
A_t=\mu(t)-\rho(t,x)z+\cdots,
\qquad
A_x=a_x(t)+O(z).
\label{eq:boundary_expansions}
\end{equation}
We adopt standard quantisation and impose the
source-free condition \(\Psi_-=0\). The boundary
condensate is identified as
\begin{equation}
{\cal O}(x,t)=\Psi_+(x,t).
\label{eq:boundary_condensate}
\end{equation}
Here \(\mu(t)\) is the chemical potential and
\(\rho(t,x)\) is the local boundary charge density.
Its spatial average,
\(\bar\rho=L_x^{-1}\int_0^{L_x}\rho(t,x)\,dx\),
is conserved by the boundary continuity equation
and periodicity, although the local density may vary
during the evolution.

The boundary sources are spatially homogeneous and
are prescribed through the driving parameter \(v(t)\):
\begin{equation}
\left.A_t\right|_{z=0}=\mu[v(t)],
\qquad
\left.A_x\right|_{z=0}=a_x(t)=-v(t)\mu[v(t)] .
\label{eq:boundary_sources}
\end{equation}
The function \(\mu(v)\) is determined by the
homogeneous stationary branch at fixed \(\bar\rho\).
During the nonlinear evolution, \(v(t)\) specifies
the imposed boundary sources and need not coincide
with the actual superfluid flow.

Writing the condensate as
\({\cal O}(x,t)=|{\cal O}(x,t)|e^{i\theta(x,t)}\),
the gauge-invariant mean superfluid flow is
\begin{equation}
\bar u_s(t)
=
\frac{2\pi N(t)/L_x-a_x(t)}{\mu(t)},
\label{eq:mean_superfluid_flow}
\end{equation}
where \(N(t)\) is the winding number defined below.
For any well-defined zero-winding configuration,
this reduces to \(\bar u_s=v(t)\). A non-zero winding
adds a quantised contribution to the mean flow.

\subsection{Winding and residence times}

The topological state of the condensate is characterised
by its integer winding number. On the periodic spatial
grid, let \({\cal O}_\ell(t)={\cal O}(x_\ell,t)\).
The phase difference across neighbouring grid points
is defined by
\begin{equation}
\delta\theta_\ell(t)
=
\arg\!\left[
{\cal O}_\ell^*(t){\cal O}_{\ell+1}(t)
\right],
\label{eq:link_phase}
\end{equation}
with periodic indexing and the principal value of
\(\arg\). The winding number is then
\begin{equation}
N(t)
=
\operatorname{round}
\left[
\frac{1}{2\pi}\sum_\ell\delta\theta_\ell(t)
\right].
\label{eq:winding_number}
\end{equation}

A phase slip changes \(N\) and is accompanied by
a local depletion of the condensate amplitude.
We retain winding measurements only when the phase
is numerically well-defined and the link-phase sum
is consistent with an integer. The detailed
reliability criteria and event-localisation procedure
are given in Appendix~\ref{app:nonlinear_checks}.

For a monotonic sequence of unit winding changes,
let \(t_{n+1\rightarrow n}\) denote the time at which
the system enters the sector \(N=n\), and
\(t_{n\rightarrow n-1}\) the time at which it leaves.
The residence time in that sector is
\begin{equation}
T_n
=
t_{n\rightarrow n-1}
-
t_{n+1\rightarrow n}.
\label{eq:sector_residence_time}
\end{equation}
Denoting the final winding by \(N_f\), the residence
time of the penultimate sector is
\begin{equation}
T_{\mathrm{pen}}=T_{N_f+1}.
\label{eq:penultimate_residence_time}
\end{equation}
Thus, \(T_{\mathrm{pen}}=T_{-4}\) for trajectories
ending at \(N_f=-5\), and \(T_{\mathrm{pen}}=T_{-5}\)
for those ending at \(N_f=-6\).

\section{Quasinormal Modes and Preparation Protocols}
\label{sec:protocols}

\subsection{QNM spectrum and mode selection}
\label{subsec:qnm}

To identify the modes relevant to the nonlinear evolution,
we follow the longitudinal quasinormal-mode (QNM) spectrum
along the homogeneous stationary branch at fixed
\(\bar\rho\). For positive \(v\), modes with negative
momentum propagate against the imposed flow. With
\(L_x=16\pi\), the allowed momenta are
\(k_m=2\pi m/L_x=m/8\), where \(m\in\mathbb Z\) is
the integer Fourier index.

We consider perturbations of the form
\begin{equation}
\delta\mathbf X_m(z)e^{-i\omega_m t+ik_m x},
\qquad
\delta\mathbf X_m
=
(\delta\psi_{\mathrm{Re}},
\delta\psi_{\mathrm{Im}},
\delta A_t,\delta A_x)_m .
\label{eq:qnm_ansatz}
\end{equation}
Linearising the bulk equations gives the generalised
eigenvalue problem
\begin{equation}
\left[
\mathsf A_{k_m}(v)+\omega_m\mathsf B_{k_m}(v)
\right]\delta\mathbf X_m=0,
\label{eq:qnm_eigenvalue}
\end{equation}
where \(\mathsf A_{k_m}\) and \(\mathsf B_{k_m}\)
are radial operators. We impose regularity at the
future horizon and vanishing perturbations of the
boundary sources. For each momentum, we follow the
longitudinal branch connected to the counter-propagating
sound mode. The eigenvalue problem and the normalisation of its physical
eigenvectors follow the procedure of Ref.~\cite{Ge:2026nyg}; the
six-mode subset used here is specified below.

With the convention \(e^{-i\omega_m t}\), the growth
rate is \(\gamma_m(v)=\operatorname{Im}\omega_m(v)\);
positive values indicate an instability. For the primary pair, we also use
\begin{equation}
\Delta\gamma_{65}(v)=\gamma_{-6}(v)-\gamma_{-5}(v).
\end{equation}
The nonlinear initial perturbation contains six selected sectors,
\(m=-3,-4,-5,-6,-7,-8\). This is the finite set used
in our calculations, not an exhaustive set of the
momenta allowed by the periodic domain.

We focus on the pair \(m=-5,-6\), corresponding to
\(k=-0.625\) and \(k=-0.750\). The amplification
velocity is chosen as
\(v_\star=0.4248833187\), close to the crossing of
their growth-rate curves. At this velocity, the two
modes have nearly equal growth rates and are the
two leading branches among the six seeded modes.
They are not assumed to be exactly degenerate.
The growth-rate curves and the selected velocities
are shown in Fig.~\ref{fig:qnm_protocol}.

\subsection{History preparation and common future}
\label{subsec:history_protocol}

We start from the homogeneous stationary background
at \(v_i=0.395\). The imposed driving parameter is
increased linearly at rate \(R=4.25\times10^{-4}\)
until it reaches \(v_h=0.4086\). It is then held
for a duration \(\tau_h\), before the ramp resumes
at the same rate and reaches
\(v_{\mathrm{obs}}=0.460\).

The choice of \(v_h\) is deliberate. It lies between
the finite-momentum instability thresholds of the
selected modes,
\(v_{\rm crit}(-0.625)\simeq0.40721\) and
\(v_{\rm crit}(-0.750)\simeq0.40972\), where \(v_{\rm crit}(k)\)
denotes the velocity at which the corresponding
growth rate changes sign. At the dwell velocity,
\(\gamma_{-5}=8.77602\times10^{-4}\), whereas
\(\gamma_{-6}=-8.42512\times10^{-4}\). Thus, during
the dwell, one member of the pair grows while the
other is damped. This gives the earlier history a
direct way to alter their relative amplitudes.

Different values of \(\tau_h\) produce different
observation times \(t_{\mathrm{obs}}\), but all
histories arrive with the same
\(v_{\mathrm{obs}}\) and \(\dot v_{\mathrm{obs}}=R\).
The ramp from \(v_h\) to \(v_{\mathrm{obs}}\) is
therefore shared by all histories. The bulk fields
are not reset at the observation point; only the
subsequent boundary driving is made identical.

From \(t_{\mathrm{obs}}\), every history follows the
same \(C^2\) quintic interpolation of duration
\(T_{\mathrm{dec}}=3\), connecting
\((v,\dot v,\ddot v)=(v_{\mathrm{obs}},R,0)\)
to \((v_\star,0,0)\). To specify the interpolation,
we introduce \(s=(t-t_{\mathrm{obs}})/T_{\mathrm{dec}}\)
and write
\begin{equation}
v(s)
=
v_{\mathrm{obs}}h_0(s)
+v_\star h_1(s)
+RT_{\mathrm{dec}}h_{\mathrm{in}}(s),
\label{eq:quintic_protocol}
\end{equation}
where \(h_0(s)=1-10s^3+15s^4-6s^5\),
\(h_1(s)=10s^3-15s^4+6s^5\), and
\(h_{\mathrm{in}}(s)=s-6s^3+8s^4-3s^5\).
These basis polynomials impose the required endpoint
values, slopes and accelerations. Since the incoming
slope is positive while \(v_\star<v_{\mathrm{obs}}\),
the interpolation contains a small common overshoot
at its beginning.

We denote the arrival time at the fixed-\(v_\star\)
stage by \(t_\star=t_{\mathrm{obs}}+T_{\mathrm{dec}}\).
The subsequent evolution is performed at constant
\(v_\star\). We refer to the identical driving from
\(t_{\mathrm{obs}}\) onwards as the common future,
and to the preceding common ramp from \(v_h\) to
\(v_{\mathrm{obs}}\) as the shared segment.

\begin{figure}[t]
    \centering
    \includegraphics[width=\textwidth]
    {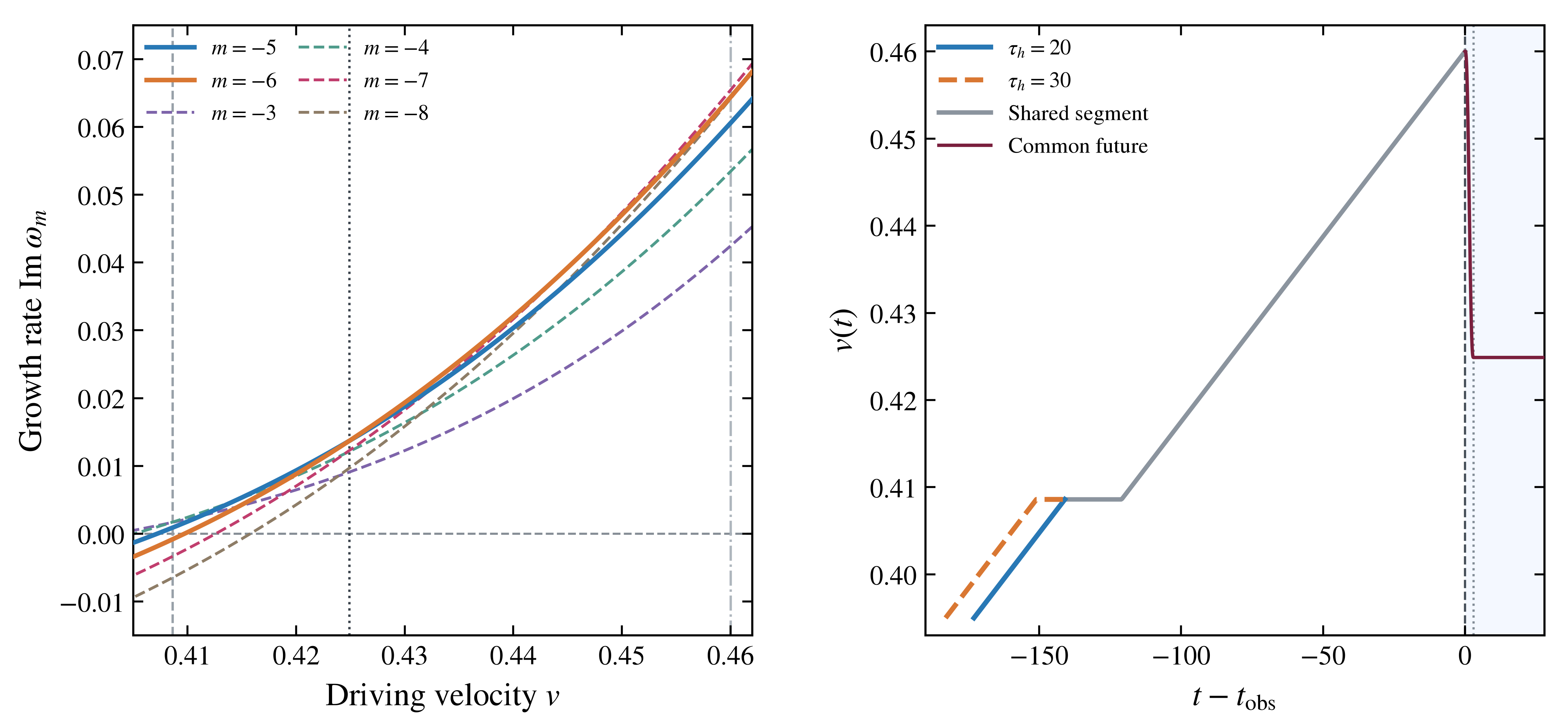}
    \caption{
    QNM growth rates and driving protocols.
    Left: the imaginary parts of the tracked QNM
    frequencies for the six seeded modes
    \(m=-3,-4,-5,-6,-7,-8\), as functions of the
    imposed driving parameter \(v\). The \(m=-5\)
    and \(m=-6\) branches are shown as solid lines,
    while the remaining four branches are dashed.
    The horizontal line marks zero growth rate.
    The vertical reference lines, from left to right,
    indicate \(v_h=0.4086\),
    \(v_\star=0.4248833187\), and
    \(v_{\mathrm{obs}}=0.460\).
    Right: representative protocols with
    \(\tau_h=20\) (blue solid) and
    \(\tau_h=30\) (orange dashed), aligned at their
    respective observation times. Their dwell
    durations differ, after which they follow the
    same ramp from \(v_h\) to \(v_{\mathrm{obs}}\)
    (grey, shared segment). From
    \(t=t_{\mathrm{obs}}\), both follow the same
    quintic interpolation and subsequent hold at
    \(v_\star\) (maroon, common future). The two
    vertical markers indicate
    \(t-t_{\mathrm{obs}}=0\) and
    \(t-t_{\mathrm{obs}}=T_{\mathrm{dec}}\).
    The shaded region marks the fixed-\(v_\star\)
    hold.
    }
    \label{fig:qnm_protocol}
\end{figure}

Figure~\ref{fig:qnm_protocol} shows two representative
histories with \(\tau_h=20\) and \(30\), aligned at
their respective observation times.

\subsection{Modal seeds and controlled preparations}
\label{subsec:modal_seeds}

The initial bulk perturbation is constructed as a
superposition of the complete QNM eigenvectors of
the six selected modes, added to the homogeneous
stationary background at \(v_i\). In the reference
preparation, each mode is normalised so that its
boundary condensate Fourier coefficient has amplitude
\(\epsilon=10^{-5}\). A fixed phase convention makes
all six initial boundary coefficients real and
positive. Thus, the perturbation is introduced in
the coupled bulk fields, rather than imposed on the
boundary condensate alone. The eigenvector normalisation follows
Ref.~\cite{Ge:2026nyg}. Here each mode is rescaled so that its
initial boundary condensate coefficient has the prescribed amplitude.

We extract the boundary Fourier coefficients from
\begin{equation}
{\cal O}_k(t)
=
\frac{1}{L_x}\int_0^{L_x}
{\cal O}(x,t)e^{-ikx}\,dx,
\qquad
A_k(t)=|{\cal O}_k(t)|.
\label{eq:boundary_fourier_amplitude}
\end{equation}
For the selected \(m=-5,-6\) pair, we define
\begin{equation}
\Delta_{56}(t)
=
\ln\!\left[
\frac{A_{-0.750}(t)}{A_{-0.625}(t)}
\right],
\qquad
\Delta_{56}^{\star}=\Delta_{56}(t_\star).
\label{eq:delta56_definition}
\end{equation}
The label \(56\) refers to the \(m=-5,-6\) pair,
with the \(m=-6\) amplitude in the numerator.
A positive value therefore favours the \(m=-6\)
component. This is a boundary Fourier-amplitude
diagnostic, not a projection onto nonlinear bulk
QNM amplitudes or a complete specification of the
bulk state.

We also use two controlled changes of the initial
modal content. In the amplitude intervention, the
initial boundary amplitudes of the \(m=-5,-6\) pair
are redistributed at fixed sum of their squared
amplitudes, while the other four seeds are left
unchanged. This fixes the boundary power of the
selected pair, not the full canonical energy of
the bulk configuration.

In the phase-only intervention, we rotate the
complete bulk QNM eigenvector of the \(m=-7\) mode
(\(k=-0.875\)) by a phase
\(e^{i\theta_{-7}(0)}\), with
\(\theta_{-7}(0)=-15^\circ,0^\circ\), or
\(+15^\circ\). The phase is applied to every
component of that eigenvector, while the other
five modes are unchanged. The phase-only comparison
uses the \(\tau_h=20\) driving protocol. All six
initial boundary amplitudes remain fixed at
\(\epsilon\).

Here \(\theta_{-7}(0)\) denotes the phase of the
initial \(m=-7\) boundary Fourier coefficient, not
the spatially dependent condensate phase
\(\theta(x,t)\) introduced in Sec.~\ref{sec:setup}.
Rotating this mode changes the relative-phase
combinations involving \(m=-7\); it does not
isolate a single independent phase invariant.
We use the term phase-only in this restricted
seed-level sense, without imposing a separate
matching condition on the full bulk energy.

\section{History-Dependent Topological Relaxation}
\label{sec:history_relaxation}

The preparations described in the previous section differ
in their earlier driving histories, but share the same
boundary driving from \(t_\star\) onwards. Their bulk
states are not reset when this common future begins.
We now ask how the memory carried by these states
appears in the subsequent nonlinear topological dynamics.

\subsection{The same topological route, different relaxation times}
\label{subsec:history_dynamics}

For every trajectory in the primary history scan, the
winding follows the sequence
\(0\rightarrow-1\rightarrow-2\rightarrow-3
\rightarrow-4\rightarrow-5\), and the final state
has \(N_f=-5\). The penultimate sector is therefore
\(N=-4\), with \(T_{\mathrm{pen}}=T_{-4}\).

\begin{figure}[t]
    \centering

    \begin{minipage}[t]{0.45\textwidth}
        \centering
        \includegraphics[width=\textwidth]
        {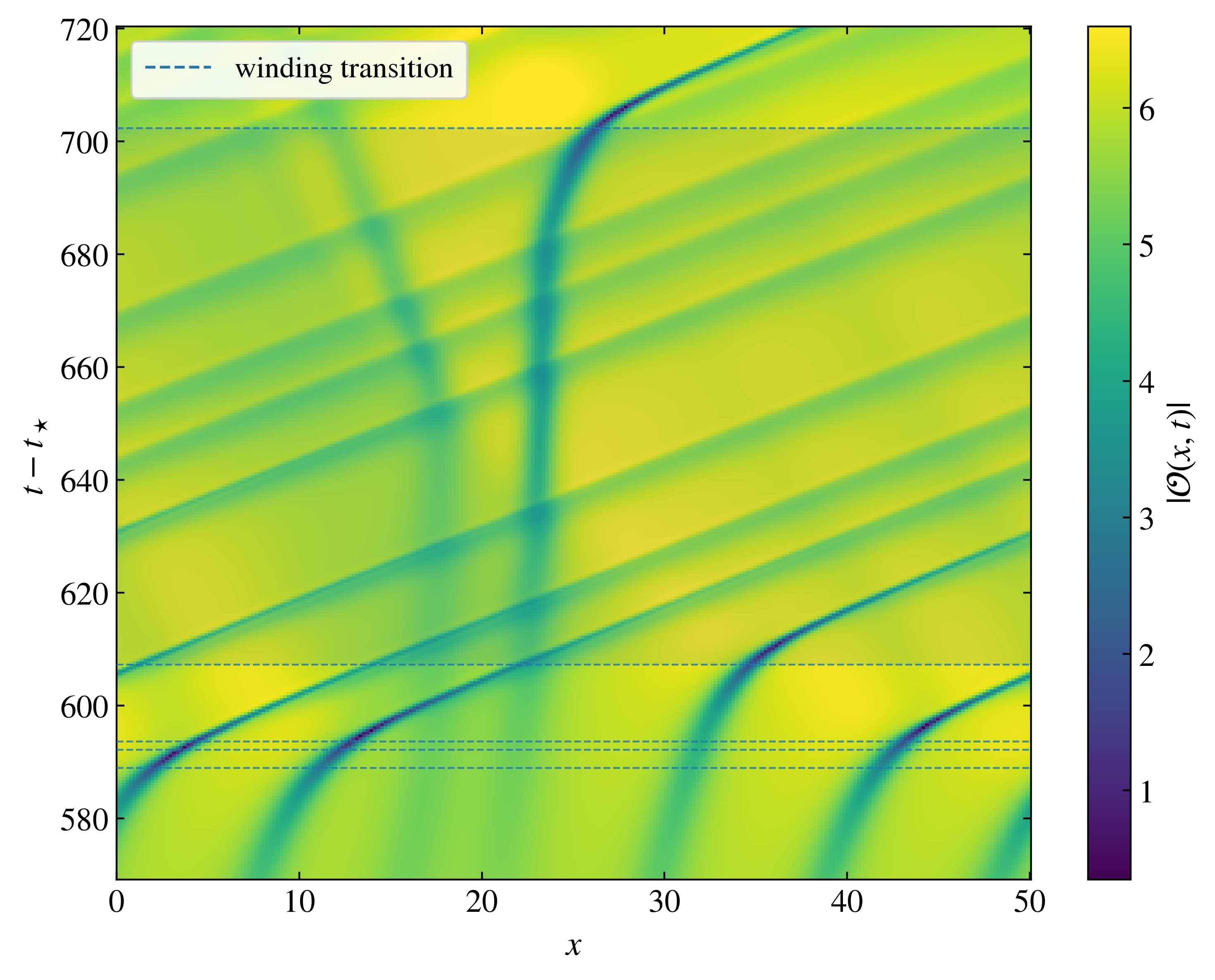}
        \textbf{(a) Condensate dynamics}
    \end{minipage}
    \hfill
    \begin{minipage}[t]{0.49\textwidth}
        \centering
        \includegraphics[width=\textwidth]
        {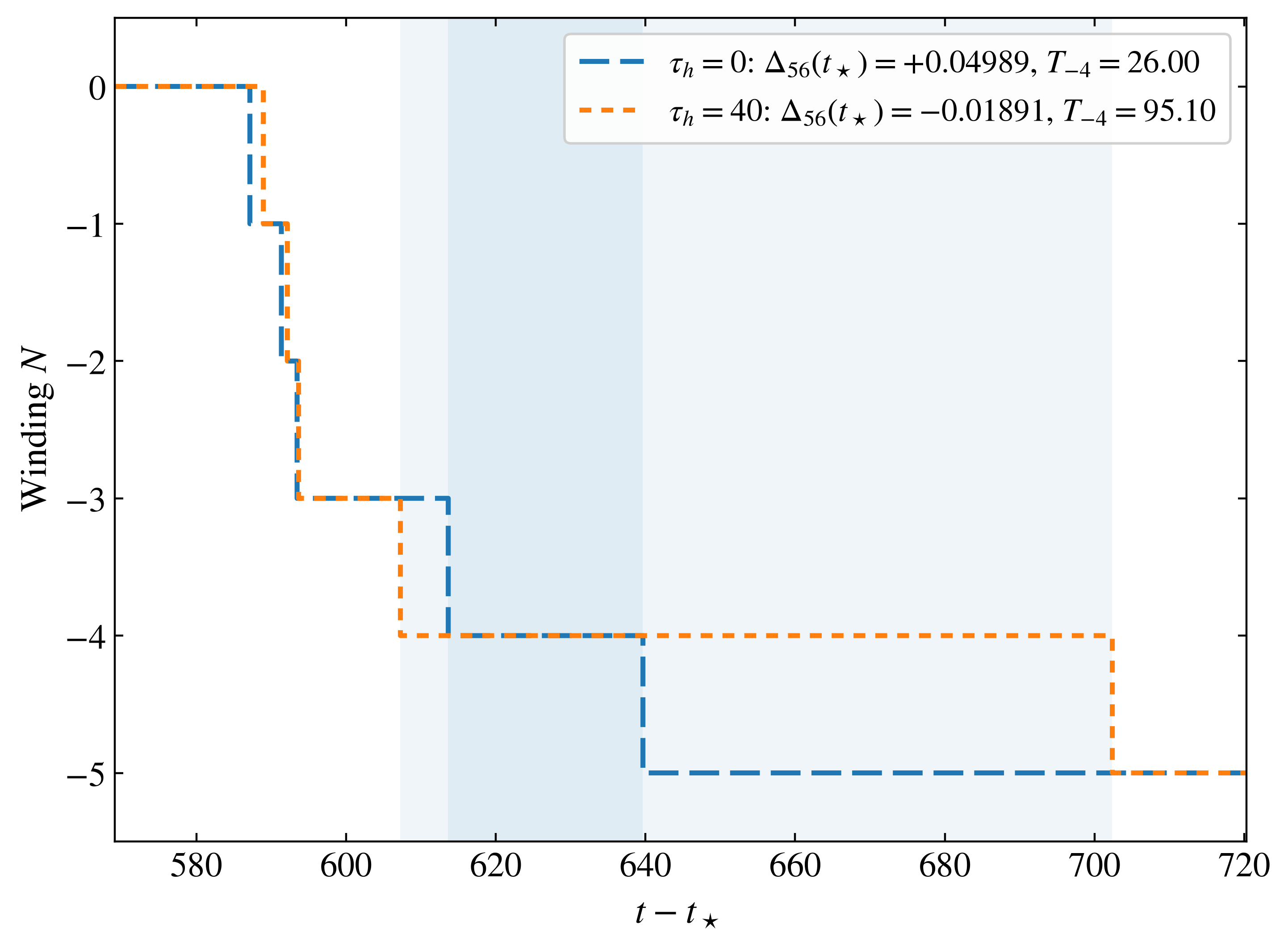}
        \textbf{(b) Winding histories}
    \end{minipage}

    \caption{
    History-dependent topological relaxation for the
    primary competing pair.
    (a) Spatiotemporal evolution of
    \(|{\cal O}(x,t)|\) for \(\tau_h=40\) at
    \(N_x=256\), with time measured from \(t_\star\).
    The horizontal dashed lines mark the five resolved
    winding transitions, while the shaded band indicates
    the residence interval in the \(N=-4\) sector.
    (b) Winding-number histories for
    \(\tau_h=0\) and \(40\), aligned at their respective
    arrival times \(t_\star\). Both trajectories reach
    \(N_f=-5\), but their residence times in the
    \(N=-4\) sector differ substantially. The shaded
    intervals mark the corresponding residence times.
    }
    \label{fig:history_relaxation}
\end{figure}

Figure~\ref{fig:history_relaxation}(a) shows the
spatiotemporal evolution of the boundary condensate
amplitude for \(\tau_h=40\). The depleted regions
are narrow and move across the periodic domain.
Each resolved winding transition is accompanied by
a local depletion of the condensate, linking the
change in winding to the underlying spatial dynamics.

The contrast between two histories is shown in
Fig.~\ref{fig:history_relaxation}(b). For
\(\tau_h=0\) and \(40\), the first
\(0\rightarrow-1\) transitions occur at
\(t-t_\star\simeq587.11\) and \(588.91\),
respectively. The first few transitions remain close
in time, but the trajectories separate more clearly
during the later stages of the evolution. In
particular, the residence time in the \(N=-4\) sector
changes from
\begin{equation}
T_{-4}(0)=26.00,
\qquad
T_{-4}(40)=95.10 .
\label{eq:history_residence_endpoints}
\end{equation}
The latter is approximately \(3.66\) times the former,
despite both trajectories reaching the same final
winding.

The difference is therefore not simply a delay of
the entire phase-slip sequence. It becomes most
pronounced in the time spent in the penultimate
sector. The earlier driving history leaves its mark
on how long the system remains in a given topological
state, even after the boundary protocol has become
identical.

The same ordering and five resolved winding changes
are retained in the \(N_x=128,192,256\) calculations
for \(\tau_h=0,20,40\). The corresponding spatial
and temporal refinement tests are collected in
Appendix~\ref{app:convergence}.

\subsection{Arrival modal imbalance within the history family}
\label{subsec:history_imbalance}

We next examine whether the difference in residence
time can be related to the modal state carried into
the common amplification stage.

For the eleven histories in the baseline scan at
\(N_x=128\), the arrival modal imbalance spans
\begin{equation}
-0.01891
\lesssim
\Delta_{56}^{\star}
\lesssim
0.04989 .
\end{equation}
Across this interval, \(T_{-4}\) decreases smoothly
from \(93.50\) to \(26.00\) as
\(\Delta_{56}^{\star}\) increases. The trend also
passes smoothly through the reversal of the modal
ordering. For example, at \(\tau_h=29\),
\begin{equation}
\Delta_{56}^{\star}=1.72\times10^{-5},
\qquad
T_{-4}=66.25 .
\end{equation}

An exponential fit describes the eleven data points
well:
\begin{equation}
T_{-4}
=
66.4203\exp\!\left(
-18.1035\,\Delta_{56}^{\star}
\right),
\label{eq:history_exponential_fit}
\end{equation}
with a coefficient of determination \(R^2=0.999350\).
We use this form only as a compact empirical description of the
finite history scan; its parameters are not assumed to define a
relation outside that preparation family.

\begin{figure}[t]
    \centering
    \includegraphics[width=0.7\textwidth]
    {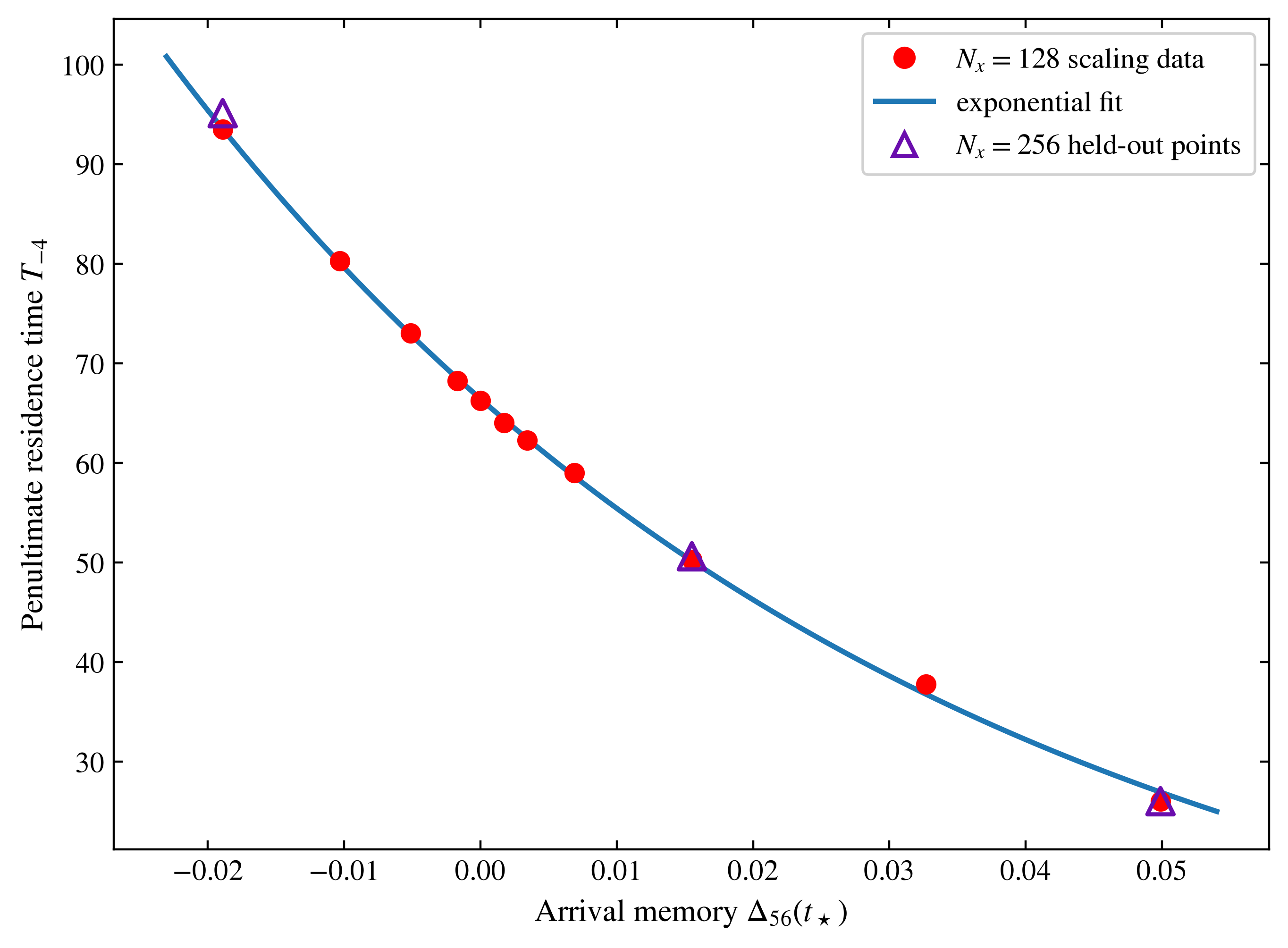}
    \caption{
    Penultimate-sector residence time \(T_{-4}\) as a
    function of the arrival modal imbalance
    \(\Delta_{56}^{\star}\) for the primary history
    family. Filled circles show the eleven
    \(N_x=128\) histories, and the solid curve is the
    exponential fit in Eq.~\eqref{eq:history_exponential_fit}.
    Open triangles show the three \(N_x=256\)
    calculations at \(\tau_h=0,20,40\), which were
    not included in the fit. All trajectories reach
    \(N_f=-5\).
    }
    \label{fig:memory_scaling}
\end{figure}

Figure~\ref{fig:memory_scaling} also shows three
\(N_x=256\) calculations at
\(\tau_h=0,20,40\). These higher-resolution points
were not used to determine the fit and remain close
to the fitted curve. They provide a check that the
observed relation is not tied to the baseline spatial
resolution.

Within this history family, the arrival modal
imbalance is therefore a useful empirical indicator
of the subsequent residence time. This correlation
should not, however, be mistaken for a complete
description of the state entering the nonlinear
evolution. Both quantities vary with the dwell
history, and their close relation within one family
does not establish that \(\Delta_{56}^{\star}\) alone
determines the later dynamics.

This leaves a more precise question than whether
driving history matters: how much of the subsequent
topological relaxation is encoded in the modal
amplitudes, and what information is carried by the
remaining structure of the initial state? We address
this by changing the modal content directly, rather
than relying on the dwell history alone.

\section{Controlled Modal Interventions: Beyond the Amplitude Imbalance}
\label{sec:controlled_interventions}

The preceding results show that the arrival modal imbalance
is closely related to the subsequent residence time within
the original history family. But this raises a further
question: does the imbalance capture the relevant information
in the prepared state, or is it only a convenient measure
of one particular way of preparing it? We address this
by changing the driving history, the modal amplitudes and
the modal phase in turn.

\subsection{Different histories with matched modal imbalance}
\label{subsec:matched_histories}

We first compare three preparations: a single dwell, a split
dwell with two separate holding stages, and a continuous
slow-window protocol without a dwell. The latter two are
constructed to match the additional integrated differential QNM
action of the single-dwell reference, relative to the common
base-rate ramp. The construction is given in
Appendix~\ref{app:matched_protocols}. We compare the three
protocols at matched arrival imbalance.

All three histories reach the same observation conditions
and subsequently follow the common future. Their arrival
modal imbalances are
\begin{equation}
\Delta_{56}^{\star}
=
0.0155129183,\quad
0.0155166544,\quad
0.0155035258,
\end{equation}
for the single-dwell, split-dwell and slow-window
preparations, respectively. The spread is only
\(1.31\times10^{-5}\). Their corresponding residence
times are
\begin{equation}
T_{-4}=50.25,\quad 50.25,\quad 50.50.
\end{equation}
All three trajectories undergo five resolved unit
winding changes and reach \(N_f=-5\).

The comparison is summarised in
Table~\ref{tab:matched_histories}. The slow-window
preparation is particularly useful: it contains no
holding stage, yet gives almost the same residence time
as the dwell-based protocols.

\begin{table}[t]
    \centering
    \begin{tabular}{lccc}
        \hline
        History
        & \(N_x\)
        & \(\Delta_{56}^{\star}\)
        & \(T_{-4}\)
        \\
        \hline
        Single dwell
        & 128
        & 0.0155129183
        & 50.25
        \\
        Split dwell
        & 128
        & 0.0155166544
        & 50.25
        \\
        Slow window
        & 128
        & 0.0155035258
        & 50.50
        \\
        \hline
        Single dwell
        & 256
        & 0.0155129156
        & 50.60118
        \\
        Slow window
        & 256
        & 0.0155035268
        & 50.50021
        \\
        \hline
    \end{tabular}
    \caption{
    Different driving histories with matched arrival modal
    imbalance. The first three rows give the production
    results at \(N_x=128\), while the last two give the
    higher-resolution comparison between the single-dwell
    and continuous slow-window preparations. All trajectories
    reach \(N_f=-5\).
    }
    \label{tab:matched_histories}
\end{table}

At \(N_x=256\), the single-dwell and slow-window arrival
imbalances differ by only \(9.39\times10^{-6}\), while
their residence times are \(50.60118\) and \(50.50021\),
respectively. The relative difference is approximately
\(0.20\%\).

The agreement is not tied to the presence of a dwell.
Within these history-based preparations, the arrival
imbalance captures an important part of the information
that determines the subsequent relaxation. It remains
to be seen whether this description survives a direct
change of the modal content.

\subsection{Redistributing the primary-pair amplitudes}
\label{subsec:amplitude_intervention}

We now change the initial amplitude distribution of the
\(m=-5,-6\) pair directly. The sum of their squared
boundary Fourier amplitudes is kept fixed, while their
relative amplitudes are varied. The other four seeded
modes retain their initial amplitudes and phases, and
the homogeneous background and subsequent driving
protocol are unchanged. Thus, the intervention redistributes
the boundary power within the primary pair; it does not
impose a matching condition on the full bulk energy.

The initial imbalance \(\Delta_{56}(0)\) is varied from
\(-0.04\) to \(+0.04\). The three principal
\(N_x=256\) results are collected in
Table~\ref{tab:controlled_amplitude}. All three
trajectories undergo five monotonic unit winding changes
and reach \(N_f=-5\).

\begin{table}[t]
    \centering
    \begin{tabular}{cccc}
        \hline
        \(\Delta_{56}(0)\)
        & \(\Delta_{56}^{\star}\)
        & \(T_{-4}\)
        & \(N_f\)
        \\
        \hline
        \(-0.04\)
        & \(-0.02448741\)
        & \(46.900261\)
        & \(-5\)
        \\
        \(0\)
        & \(+0.01551292\)
        & \(50.601180\)
        & \(-5\)
        \\
        \(+0.04\)
        & \(+0.05551328\)
        & \(55.799149\)
        & \(-5\)
        \\
        \hline
    \end{tabular}
    \caption{
    Controlled redistribution of the initial boundary
    amplitudes of the \(m=-5,-6\) pair at \(N_x=256\).
    The sum of the squared amplitudes of the primary pair
    is fixed, while the four spectator-mode seeds are
    unchanged.
    }
    \label{tab:controlled_amplitude}
\end{table}

The trend is opposite to that of the original history
family. Here, increasing the arrival imbalance increases
the residence time, whereas the history-family fit
decreases with \(\Delta_{56}^{\star}\). The amplitude
intervention is a different direction through the space
of initial states, so the history-family fit cannot be
read as a universal map from modal imbalance to residence
time.

The exponential relation found in Sec.~\ref{sec:history_relaxation}
is therefore specific to the original history-based
preparations. It cannot be used as a universal mapping
from the arrival amplitude imbalance to the subsequent
residence time.

\subsection{Phase sensitivity at matched modal amplitudes}
\label{subsec:phase_intervention}

The amplitude intervention shows that the arrival
imbalance is not sufficient to characterise the prepared
state. We next ask whether information carried by the
modal phases also matters.

We use the \(\tau_h=20\) preparation and rotate the
complete bulk QNM eigenvector of the \(m=-7\) mode by
\(\theta_{-7}(0)=-15^\circ\) or \(+15^\circ\). The other
five eigenvectors are left unchanged. All six initial
boundary Fourier amplitudes retain the same magnitudes,
as do the primary-pair power, homogeneous background
and subsequent driving protocol.

The two preparations arrive with almost identical
\(\Delta_{56}^{\star}\), differing by less than
\(10^{-9}\). Their event-refined residence times,
however, are
\begin{equation}
T_{-4}(-15^\circ)=46.175881,
\qquad
T_{-4}(+15^\circ)=56.001098.
\label{eq:phase_only_residence}
\end{equation}
Both trajectories undergo five monotonic unit winding
changes and reach \(N_f=-5\). The results are summarised
in Table~\ref{tab:phase_only_intervention}.

\begin{table}[t]
    \centering
    \begin{tabular}{cccc}
        \hline
        \(\theta_{-7}(0)\)
        & \(\Delta_{56}^{\star}\)
        & \(T_{-4}\)
        & \(N_f\)
        \\
        \hline
        \(-15^\circ\)
        & \(0.01551291535\)
        & \(46.175881\)
        & \(-5\)
        \\
        \(+15^\circ\)
        & \(0.01551291617\)
        & \(56.001098\)
        & \(-5\)
        \\
        \hline
    \end{tabular}
    \caption{
    Phase-only intervention obtained by rotating the complete
    \(m=-7\) bulk QNM eigenvector, at \(N_z=32\) and
    \(N_x=256\). The initial boundary modal amplitudes are
    unchanged, and the arrival imbalances differ by less
    than \(10^{-9}\). The quoted residence times use the
    refined event-localisation data.
    }
    \label{tab:phase_only_intervention}
\end{table}

The difference is
\begin{equation}
\Delta T_{-4}=9.825216,
\end{equation}
corresponding to a symmetric relative difference of
approximately \(19.23\%\). The response retains its sign
under the \(N_z=28\) to \(32\) radial comparison and
when the Runge--Kutta step is halved. Refining the event
localisation gives a residence-time bracket bound of
approximately \(0.0264\). These are numerical stability
checks rather than a continuum extrapolation. The radial
and event-localisation tests are given in
Appendix~\ref{app:phase_radial_convergence} and
Appendix~\ref{app:phase_event_time_refinement}.

The individual transition times also show that the
difference is not a uniform displacement of the entire
phase-slip sequence. The first \(0\rightarrow-1\)
transition occurs earlier for the \(+15^\circ\)
preparation, at \(t-t_\star\simeq585.48\), compared
with \(590.73\) for the \(-15^\circ\) preparation.
Yet the final \(-4\rightarrow-5\) transition occurs
later, at approximately \(668.43\) rather than \(654.73\).
The phase intervention therefore changes how the
relaxation time is distributed among successive
topological transitions.

\subsection{What the amplitude imbalance leaves out}
\label{subsec:beyond_amplitude_imbalance}

The three comparisons give a more precise picture of
the information retained by the prepared state.
Different histories with matched arrival imbalance
produce nearly identical residence times. Direct
amplitude redistribution, however, reverses the trend
found in the original history family. Most strikingly,
the phase-only intervention produces substantially
different residence times at essentially the same
arrival imbalance.

The arrival amplitude ratio is therefore useful, but
not sufficient, to describe the subsequent topological
relaxation across different preparation families.
At least some of the information relevant to the later
dynamics lies in the complex multimode state, beyond
the individual mode amplitudes. The phase intervention
shows this directly, although it does not isolate a
single independent phase invariant: rotating the
\(m=-7\) eigenvector changes all relative-phase
combinations involving that mode.

We now turn to the local phase-slip dynamics to examine
how these differences in the prepared state appear
in the evolving condensate.

\section{Local Phase-Slip Structure and Multimode Maxwell Feedback}
\label{sec:local_phase_slip}

The phase dependence found in the previous section also
appears in the local evolution of the condensate. We now
look more closely at the final phase-slip event and at
the channels through which the prepared modal phases can
enter the nonlinear dynamics.

\subsection{Local structure of the phase slip}
\label{subsec:local_slip_structure}

We compare the three preparations with
\(\theta_{-7}(0)=-15^\circ,0^\circ,+15^\circ\),
using the same \(\tau_h=20\) protocol. Our attention
is on the final transition \(N=-4\rightarrow-5\). We denote
the spatial and temporal centres returned by the independent
phase-slip diagnostic by \(x_{\rm slip}\) and \(t_{\rm slip}\);
the latter is the midpoint of the final event bracket.

\begin{figure}[t]
    \centering
    \includegraphics[width=\textwidth]
    {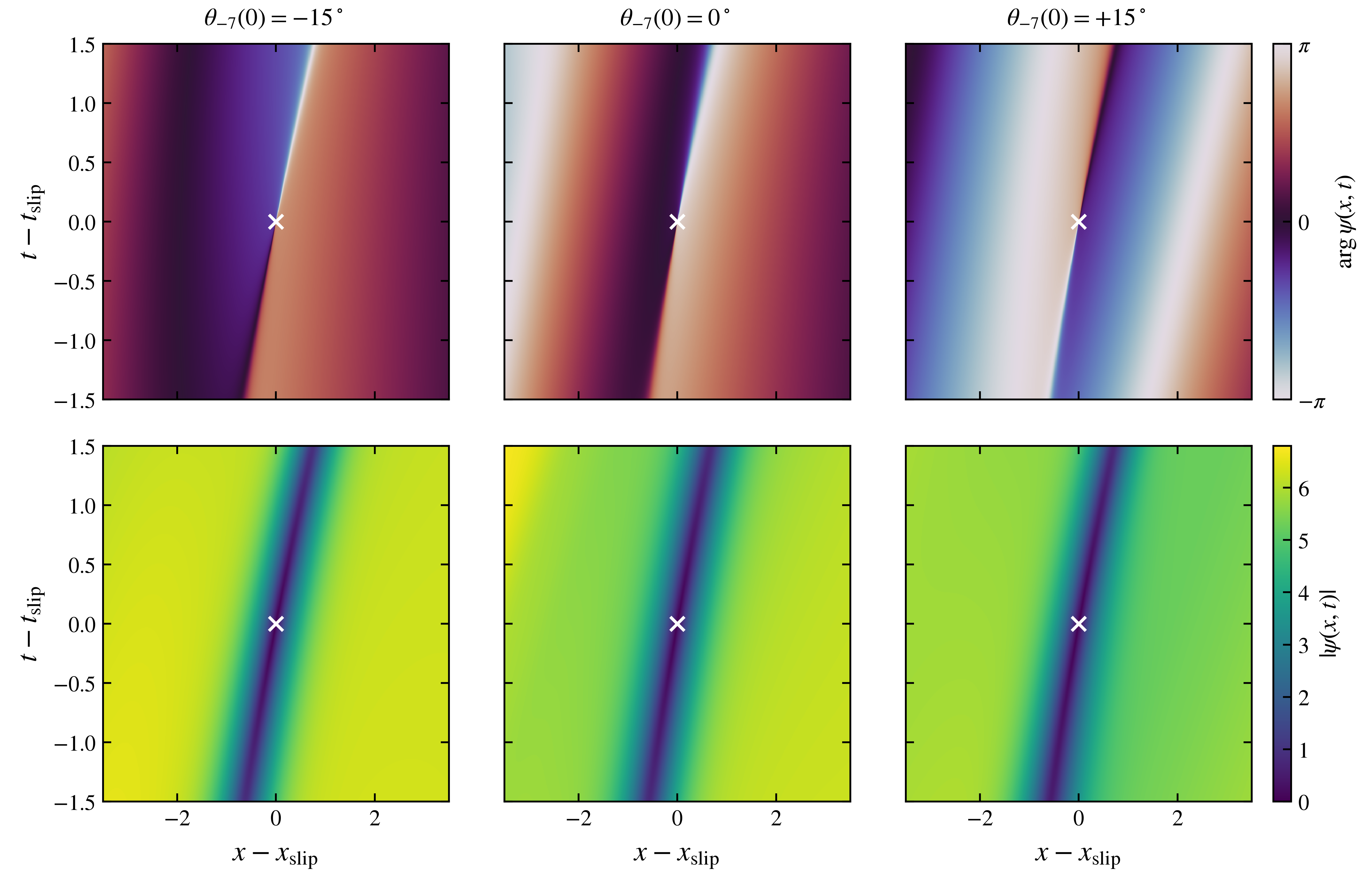}
    \caption{
    Local space-time structure of the final phase-slip
    event for three initial phases of the \(m=-7\) mode.
    The columns correspond to
    \(\theta_{-7}(0)=-15^\circ,0^\circ,+15^\circ\).
    The upper row shows the phase of the boundary
    condensate \({\cal O}(x,t)\), and the lower row
    its amplitude. The coordinates are centred on
    the corresponding event,
    \(x-x_{\mathrm{slip}}\) and
    \(t-t_{\mathrm{slip}}\); the white cross marks
    the event centre identified by the independent
    phase-slip diagnostics. The phase colour scale is
    periodic, with \(-\pi\) and \(+\pi\) representing
    the same phase. The amplitude panels share a
    common colour scale. The local replay uses
    \(N_x=8192\). The limitations of these maps are discussed
    in Appendix~\ref{app:local_slip_structure}.
    }
    \label{fig:local_phase_slip}
\end{figure}

Figure~\ref{fig:local_phase_slip} shows the complex
boundary condensate in a neighbourhood of this event.
The coordinates are centred on the event position and
time, \(x-x_{\mathrm{slip}}\) and
\(t-t_{\mathrm{slip}}\). The upper row displays the
phase, while the lower row shows the condensate
amplitude. The winding diagnostic records a unit change for each
transition, while the amplitude maps show local condensate
depletion in the event region. We do not assign a plaquette
charge to these holographic maps. The three final events occur
near the same position,
\(x_{\mathrm{slip}}\simeq27\). Thus, changing the
initial modal phase does not move the final slip to
a different part of the ring; the differences lie
in the evolution of the local condensate structure.

To compare the recovery of the condensate between
preparations, we introduce the global notch diagnostic
\begin{equation}
r(t)
=
\frac{\min_x|{\cal O}(x,t)|}
{\operatorname{median}_x|{\cal O}(x,t)|}.
\label{eq:condensate_notch_ratio}
\end{equation}
Let \(t_{0.5}^{\mathrm{last}}\) be the last downward
crossing of \(r=0.5\) within the \(N=-4\) residence
interval. We define the corresponding delay by
\begin{equation}
\tau_{0.5}
=
t_{0.5}^{\mathrm{last}}-t_{-3\rightarrow-4}.
\label{eq:notch_redevelopment_delay}
\end{equation}
The results are
\begin{equation}
\tau_{0.5}
\simeq
38.15,\quad42.38,\quad45.12,
\end{equation}
for the initial phases \(-15^\circ,0^\circ,+15^\circ\),
respectively.

The ordering follows that of the residence times
\(T_{-4}\): the \(+15^\circ\) preparation takes longer
to develop a deep global notch than the \(-15^\circ\)
preparation. This is a useful local diagnostic of the
different relaxation histories, although it does not
identify or track a particular defect throughout the
evolution. The threshold \(r=0.5\) is operational,
not a universal phase-slip criterion.

\subsection{Phase-sensitive multimode interference}
\label{subsec:multimode_interference}

The role of the modal phases can be seen explicitly
in the Fourier structure of the condensate density.
At the arrival time \(t_\star\), write
\(c_m={\cal O}_{k_m}(t_\star)
=A_{k_m}(t_\star)e^{i\theta_m}\), where
\(\theta_m\) is the phase of the corresponding
boundary Fourier coefficient. The spacing between
neighbouring modes is
\(\kappa=2\pi/L_x=1/8\). The contribution of the
selected modes \(m=-5,-6,-7\) to the \(+\kappa\)
harmonic of \(|{\cal O}|^2\) is
\begin{equation}
B_\kappa^{(567)}
=
c_{-5}c_{-6}^*
+
c_{-6}c_{-7}^* .
\label{eq:triad_density_harmonic}
\end{equation}
Its phase dependence is governed by
\begin{equation}
\Phi_{567}
=
\arg\!\left[c_{-5}c_{-7}(c_{-6}^*)^2\right]
=
\theta_{-5}+\theta_{-7}-2\theta_{-6},
\label{eq:triad_phase}
\end{equation}
and the following identity holds exactly:
\begin{equation}
|B_\kappa^{(567)}|^2
=
A_{k_{-6}}^2
\left[
A_{k_{-5}}^2+A_{k_{-7}}^2
+2A_{k_{-5}}A_{k_{-7}}\cos\Phi_{567}
\right].
\label{eq:triad_interference}
\end{equation}

Since \(k_{-5}+k_{-7}-2k_{-6}=0\), the phase
combination \(\Phi_{567}\) is unchanged by a rigid
spatial translation. It is also unchanged by a common
constant phase shift in the fixed boundary gauge,
although it is not invariant under arbitrary local
gauge transformations.

For the phase-only preparations, the magnitude of
this selected three-mode contribution increases by
approximately \(6.89\%\) between the
\(-15^\circ\) and \(+15^\circ\) cases. The full
\(+\kappa\) harmonic of \(|{\cal O}|^2\), however,
does not follow the same monotonic trend. Other
Fourier modes contribute to that harmonic and can
partly cancel the selected contribution.

The identity therefore provides an explicit example
of how relative phase enters the condensate density.
It does not close the nonlinear dynamics into a
three-mode system, nor does it establish
\(\Phi_{567}\) as the variable controlling the
residence time. The full multimode structure must
remain part of the physical picture.

\subsection{Maxwell feedback and its limitations}
\label{subsec:maxwell_feedback}

The bulk equations provide a natural route for this
phase information to enter the gauge-field dynamics.
Varying the matter action gives
\begin{equation}
\nabla_\mu F^{\mu\nu}=J^\nu,
\qquad
J^\nu
=
i\left[
\Psi^*D^\nu\Psi
-
\Psi(D^\nu\Psi)^*
\right],
\label{eq:maxwell_scalar_current}
\end{equation}
where \(J^\nu\) is the bulk scalar current. Its
bilinear structure allows different scalar modes
to contribute through their relative phases. The resulting gauge-field response feeds back into the
scalar equation through the gauge-covariant derivative
and field-strength terms.

At fixed gauge field, the scalar equation contains
no explicit local quartic self-interaction. The
nonlinear coupling considered here instead arises
through the coupled evolution of the scalar and
gauge fields.

Appendix~\ref{app:maxwell_feedback} decomposes the
\(q=+\kappa\) component of the \(A_x\) equation into radial,
bare-current, gauge-density and electric terms, and shows how
the selected three-mode phase combination enters this coupling.
The decomposition identifies an allowed feedback channel, but
it does not establish a dominant source term or a common
threshold preceding the final phase slip.

Taken together, the winding changes, the local condensate
depletion and the Maxwell coupling analysis are consistent
with a multimode Maxwell--scalar feedback process.
They do not identify a single dominant Fourier
triad or source term as the cause of the lifetime
difference. The phase-sensitive response belongs
to the coupled nonlinear evolution as a whole,
rather than to one isolated modal threshold.

We next examine how far this response persists when
the common future and the background are changed.

\section{Robustness and Generality}
\label{sec:robustness_generality}

The results above were obtained near one particular crossing of the QNM growth-rate curves and on a single charge-density background.
We now ask how much of the phase-sensitive response
survives when the common future is changed, and when
the same intervention is repeated on a different
background. These tests also reveal an important
feature of the relaxation: the phase response need
not have the same effect on successive winding sectors.

\subsection{Changing the common future}
\label{subsec:common_future}

We first return to the original background,
\(\bar{\rho}=10.0417\), and vary the constant velocity
reached after a three-unit taper. We denote this
common-future velocity by \(v_c\), to distinguish it
from the earlier history-writing velocity \(v_h\).
The phase intervention is unchanged: the complete
bulk QNM eigenvector of the \(m=-7\) mode is rotated
by \(\theta_{-7}(0)=\pm15^\circ\), while the other
five modes retain their original amplitudes and phases.

For each \(v_c\), the two preparations use the same
\(\tau_h=20\) history and the same subsequent quintic taper,
with the target \(v_\star\) in Eq.~\eqref{eq:quintic_protocol}
replaced by \(v_c\). We denote the endpoint of this taper by
\(t_c=t_{\mathrm{obs}}+T_{\mathrm{dec}}\), and the corresponding
arrival imbalance by \(\Delta_{56}^{(c)}=\Delta_{56}(t_c)\).
The two phase preparations have closely matched
\(\Delta_{56}^{(c)}\) at each \(v_c\); when \(v_c=v_\star\),
\(t_c=t_\star\). To compare the spectral ordering with the
nonlinear response, we use \(\Delta\gamma_{65}(v)\) defined
in Sec.~\ref{subsec:qnm}, and measure the phase contrast through
\(\Delta T_{\mathrm{pen}}
=T_{\mathrm{pen}}(+15^\circ)-T_{\mathrm{pen}}(-15^\circ)\).
We write \(N_{\mathrm{pen}}=N_f+1\) for the winding sector
whose residence time is \(T_{\mathrm{pen}}\).

The \(N_x=256\) results are collected in
Table~\ref{tab:common_future_phase_response}.

\begin{table}[t]
    \centering
    \scriptsize
    \setlength{\tabcolsep}{2.5pt}
    \begin{tabular}{cccccc}
        \hline
        \(v_c\)
        & \(\Delta\gamma_{65}\)
        & \((N_f,N_{\mathrm{pen}})\)
        & \shortstack{\(T_{\mathrm{pen}}\)\\\(-15^\circ\)}
        & \shortstack{\(T_{\mathrm{pen}}\)\\\(+15^\circ\)}
        & \(\Delta T_{\mathrm{pen}}\)
        \\
        \hline
        \(0.420000\)
        & \(-5.334\times10^{-4}\)
        & \((-5,-4)\)
        & \(31.201\)
        & \(34.898\)
        & \(+3.697\)
        \\
        \(0.424883\)
        & \(0\)
        & \((-5,-4)\)
        & \(46.198\)
        & \(56.001\)
        & \(+9.803\)
        \\
        \(0.430000\)
        & \(+5.680\times10^{-4}\)
        & \((-6,-5)\)
        & \(4.899\)
        & \(1.901\)
        & \(-2.998\)
        \\
        \(0.440000\)
        & \(+1.685\times10^{-3}\)
        & \((-6,-5)\)
        & \(2.502\)
        & \(1.901\)
        & \(-0.601\)
        \\
        \hline
    \end{tabular}
    \caption{
    Phase-dependent penultimate-sector residence times
    for four common-future velocities on the original
    background. All calculations use \(N_x=256\) and
    \(N_z=28\). The growth-rate difference is defined
    for the original \((-5,-6)\) pair, even when that
    pair no longer occupies the two leading positions.
    The \(v_c=0.424883\) row is the independent
    near-degeneracy phase-only reference.
    All times are in simulation units.
    }
    \label{tab:common_future_phase_response}
\end{table}

The phase response changes as the common future is
varied. At \(v_c=0.420\) and at the original
near-degeneracy, the \(+15^\circ\) preparation spends
longer in the penultimate sector. At the two higher
velocities, the ordering is reversed.

There is, however, an important distinction between
these cases. For the first two velocities, the
trajectories reach \(N_f=-5\), so that
\(N_{\mathrm{pen}}=-4\). At \(v_c=0.430\) and
\(0.440\), they instead reach \(N_f=-6\), and the
penultimate sector is \(N=-5\). The change in the
sign of \(\Delta T_{\mathrm{pen}}\) therefore does
not represent a reversal of the response of one
and the same winding sector.

Indeed, the first completed residence in \(N=-4\)
remains longer for the \(+15^\circ\) preparation.
At \(v_c=0.430\), it increases from \(26.903\) to
\(44.701\), while at \(v_c=0.440\) it increases
from \(10.699\) to \(12.898\). In both cases, the
subsequent residence in \(N=-5\) is shorter for
the \(+15^\circ\) preparation.

The phase intervention thus redistributes the time
spent in successive winding sectors. It does not
simply accelerate or delay the entire sequence of
phase slips.

The values in the table are taken from the
high-resolution monotonic trajectories. In the
\(N_x=128\) exploratory runs, some apparent
back-slips disappear at \(N_x=256\), so the
lower-resolution event histories are not used to
define the residence times. The corresponding
refinement checks are given in
Appendix~\ref{app:common_future_validation}.

It is also worth noting that changing \(v_c\)
changes the stationary background, the absolute
QNM growth rates and the subsequent nonlinear
evolution, not merely the difference between the
two primary growth rates. The observed response
should therefore not be regarded as a function of
\(\Delta\gamma_{65}\) alone.

For the near-degeneracy row, the production
\(N_z=28\) value is used here. The later
\(N_z=32\) event-bracket refinement quoted in
Sec.~\ref{subsec:phase_intervention} gives
\(T_{-4}(-15^\circ)=46.175881\), rather than
\(46.197832\). The shift lies within the original
event bracket and does not affect the phase
comparison.

\subsection{When the leading competing pair changes}
\label{subsec:outside_pair_window}

The \(v_c=0.440\) case provides a further test of
the role of the original competing pair. Within
the six seeded modes, the growth rates are ordered
as
\begin{equation}
\gamma_{-6}>\gamma_{-7}>\gamma_{-5}.
\end{equation}
In particular,
\(\gamma_{-6}-\gamma_{-7}
=3.7163\times10^{-4}\), while
\(\gamma_{-7}-\gamma_{-5}
=1.3131\times10^{-3}\).
The original \((-5,-6)\) pair is therefore no
longer the leading pair; the two fastest seeded
modes are now \((-6,-7)\).

The phase-rotated \(m=-7\) mode has consequently
entered the leading pair. Nevertheless, the
phase-sensitive response remains. Both
\(N_x=256\) trajectories undergo six monotonic
unit winding changes and reach the same observed
final winding, \(N_f=-6\). Their residence times
in the penultimate \(N=-5\) sector are \(2.502\)
and \(1.901\) for the \(-15^\circ\) and
\(+15^\circ\) preparations, respectively.

The first residence in \(N=-4\), by contrast,
increases from \(10.699\) to \(12.898\).
The difference between these two residence-time
responses reinforces the point made above: the
initial phase affects how the relaxation unfolds
across successive sectors, rather than setting
a single clock for the entire evolution.

This result extends the phase-sensitive response
beyond the original \((-5,-6)\) competing-pair
window. It does not, however, establish a
single-mode-dominant limit. The \(m=-7\) mode
has become the second-fastest seeded mode, and
the calculation still involves all six modes.
The spatial-refinement results are collected in
Appendix~\ref{app:common_future_validation}.

\subsection{A second charge-density background}
\label{subsec:second_density}

We finally repeat the phase-only intervention on
a second background with mean charge density
\(\bar{\rho}_2=10.543785\), compared with
\(\bar{\rho}=10.0417\) in the original case.
The new background has its own near-degenerate
primary pair at
\(v_\star^{(2)}=0.4383063373\).

The independently checked QNM spectrum gives
\(\gamma_{-6}-\gamma_{-5}
=1.039\times10^{-11}\), with the six seeded modes
ordered as \((-6,-5,-7,-4,-8,-3)\).
The next mode, \(m=-7\), lies below the leading
pair by approximately \(1.264\times10^{-3}\).

To prepare this background, the original
velocity protocol is shifted by
\(0.0134230186\), while the ramp rate, dwell
duration and deceleration time are retained.
This places the new background near its own
QNM crossing. It is a retuned preparation,
rather than a change of charge density alone.

We again rotate the complete bulk QNM eigenvector
of \(m=-7\) by \(\pm15^\circ\), leaving the
magnitudes of all six initial boundary Fourier
amplitudes unchanged. Let \(t_{\mathrm{obs}}^{(2)}\) be the time at which the shifted
history reaches \(v_{\mathrm{obs}}^{(2)}\). The shifted taper ends at
\(t_\star^{(2)}=t_{\mathrm{obs}}^{(2)}+T_{\mathrm{dec}}\).
We write \(\Delta_{56}^{\star(2)}=\Delta_{56}(t_\star^{(2)})\)
and \(\Phi_{567}^{\star(2)}=\Phi_{567}(t_\star^{(2)})\).
The two preparations arrive with almost identical
\(\Delta_{56}^{\star(2)}\), but their
\(\Phi_{567}^{\star(2)}\) values differ by approximately
\(30^\circ\). The resulting \(N_x=256\) evolutions are
summarised in
Table~\ref{tab:second_density_phase_response}.

\begin{table}[t]
    \centering
    \small
    \setlength{\tabcolsep}{4pt}
    \begin{tabular}{ccccc}
        \hline
        \(\theta_{-7}(0)\)
        & \(\Delta_{56}^{\star(2)}\)
        & \(\Phi_{567}^{\star(2)}\)
        & \(T_{\mathrm{pen}}\)
        & \(N_f\)
        \\
        \hline
        \(-15^\circ\)
        & \(0.022570076004\)
        & \(-42.6820^\circ\)
        & \(39.402\)
        & \(-5\)
        \\
        \(+15^\circ\)
        & \(0.022570075980\)
        & \(-12.6821^\circ\)
        & \(47.603\)
        & \(-5\)
        \\
        \hline
    \end{tabular}
    \caption{
    Phase-only intervention on the second background,
    \(\bar{\rho}_2=10.543785\), at
    \(N_x=256\) and \(N_z=28\). The initial boundary
    modal amplitudes are unchanged between the two
    preparations. Both trajectories reach the same
    observed final winding through five monotonic
    unit changes. Since \(N_f=-5\), the penultimate
    sector is \(N=-4\), and \(T_{\mathrm{pen}}=T_{-4}\).
    }
    \label{tab:second_density_phase_response}
\end{table}

The two values of \(\Delta_{56}^{\star(2)}\) differ by only
\(2.40\times10^{-11}\), whereas the residence times differ by
\begin{equation}
\Delta T_{-4}
=
T_{-4}(+15^\circ)-T_{-4}(-15^\circ)
=
8.200851.
\end{equation}
The corresponding event-bracket difference lies
between \(7.998902\) and \(8.402799\), so its
sign is unaffected by the event-localisation
uncertainty.

The response also survives spatial refinement.
Between \(N_x=128\) and \(N_x=256\), the
individual residence times change by approximately
\(0.515\%\) and \(1.715\%\) for the
\(-15^\circ\) and \(+15^\circ\) preparations,
respectively. Both resolutions give the same
ordering of the phase response, and the
high-resolution trajectories have no detected
back-slips or non-unit winding changes. This is
a two-resolution stability check, not a formal
continuum extrapolation. The complete comparison
is given in Appendix~\ref{app:second_density_validation}.

The difference is also comparable in size to
the phase response on the original background.
The effect therefore survives a change of
background and a corresponding retuning of the
driving protocol.

There is one qualification. The arrival imbalance
on the second background is approximately
\(0.02257\), rather than \(0.01551\) in the
original preparation. We have therefore established
the persistence of phase sensitivity in a second
retuned background, not a density-independent
relation at fixed \(\Delta_{56}^{\star}\).

The results above show that phase-sensitive
topological relaxation is not confined to one
common-future velocity or to the original
charge-density background. They also reveal that
its effect is distributed across successive
winding sectors and depends on the evolving
multimode state. These calculations do not
establish universality over arbitrary backgrounds,
driving histories or modal spectra. In particular,
we have not reached a genuinely single-mode-dominant
regime, and the final winding numbers remain
numerically observed plateaux rather than proofs
of infinite-time attractors.

Within these limits, the central conclusion is
that the complex modal state continues to shape
topological relaxation even after the driving
protocol has been changed or the background
has been retuned.

\section{Conclusion and Discussion}
\label{sec:conclusion}

We have studied how the complex modal state prepared by an
earlier drive shapes the subsequent topological relaxation
of a holographic superfluid. The main result is that the
final winding and the time taken to reach it need not
carry the same information. Different preparations can
lead to the same observed final winding under an identical
future drive, yet spend markedly different times in the
intermediate topological sectors.

For the primary competing pair, all histories in the
original scan reach $N_f=-5$, while the residence time
in the penultimate $N=-4$ sector varies by a factor of
approximately $3.6$. Within this history family, the
arrival modal imbalance $\Delta_{56}^{\star}$ provides
a useful empirical description of the variation. The
matched-history tests give this relation a more physical
footing: a single dwell, a split dwell and a continuous
slow-window preparation produce nearly the same subsequent
relaxation when their arrival imbalances are matched.
In particular, the single-dwell and slow-window results
agree to within $0.2\%$ in the refined comparison, despite
the absence of a holding stage in the latter. The relevant
information is therefore not simply the duration of a
dwell.

The controlled interventions reveal the limits of this
description. Redistributing the amplitudes of the primary
mode pair reverses the trend found in the original
history family. More decisively, rotating the complete
bulk eigenvector of the $m=-7$ mode changes the
residence time while leaving the initial boundary modal
amplitudes unchanged and keeping the arrival imbalance
matched to numerical precision. The refined comparison
gives a difference of approximately $9.83$ in
$T_{-4}$, corresponding to a symmetric relative
difference of about $19.23\%$. A single amplitude
ratio is therefore not sufficient to describe the
subsequent nonlinear evolution across different
preparations.

Changing the common future reveals a further feature
of the response. A phase intervention can lengthen
the residence in one winding sector while shortening
that in the next. At $v_c=0.430$, for example,
changing the initial phase from $-15^\circ$ to
$+15^\circ$ increases $T_{-4}$ from $26.903$
to $44.701$, but decreases $T_{-5}$ from $4.899$
to $1.901$. The effect is thus not a uniform delay
of the phase-slip sequence. It changes how the
relaxation time is distributed among successive
topological sectors. A related response persists at
$v_c=0.440$, where the leading seeded QNM pair has
changed, and on a second charge-density background
with $\bar{\rho}=10.543785$. These tests establish
robustness over the backgrounds and driving protocols
examined here, rather than universality over arbitrary
superflows or modal spectra.

The local evolution provides a spatial counterpart
to this timing dependence. Winding changes are
accompanied by local condensate depletion and a rearrangement
of the phase field around the event region. The Fourier
analysis shows explicitly how relative modal phases enter
the condensate density, while the Maxwell equation allows
the resulting phase-sensitive currents to feed back
into the scalar dynamics. The selected three-mode
contribution identifies one such channel, but neither
$\Phi_{567}$ nor any individual Maxwell source term
has been established as the unique mediator. The
evidence points instead to the coupled evolution of
the complex multimode state.

An independent one-dimensional dissipative
Gross--Pitaevskii model provides a further example of
phase-dependent winding-sector relaxation. We compare
two initial states that differ only in the phase coordinate
$\Phi_{567}$, while their initial grand-potential densities,
seeded particle- and hole-mode powers, mean densities
and momentum densities are matched to numerical
precision. Both states undergo the same six unit winding
changes and reach the same observed final sector,
$N_f=-6$, but their residence times in the $N=-5$
sector differ by a factor of approximately $3.11$.
The model contains no Maxwell field and is not a
reduction of the holographic theory; it shows that
phase-dependent winding-sector lifetimes can also arise
in a distinct dissipative superfluid model. The
construction and numerical checks are given in
Appendix~\ref{app:dgpe_bridge}.

There is a natural connection to real superfluids.
In a ring, the winding number labels quantised
circulation,
\begin{equation}
\oint \mathbf{v}_{\mathrm{s}}\cdot d\boldsymbol{\ell}
=
N\frac{h}{M_{\mathrm{at}}},
\end{equation}
where $\mathbf{v}_{\mathrm{s}}$ is the superfluid
velocity, $M_{\mathrm{at}}$ is the atomic mass
and $h$ is Planck's constant. Phase slips connect
distinct persistent-current sectors
\cite{Varoquaux2015}. The residence time in a given
sector therefore has a direct counterpart in the
waiting time before the circulation changes.

A possible experimental test would use a ring-shaped
atomic superfluid in which a small number of azimuthal
modes are prepared with matched powers but different
relative phases. The exact mode labels and phase
combination need not be those used in our calculations;
the essential requirement is to distinguish preparations
that have similar modal populations but different
coherent phase structure. After preparation, the same
rotational drive or weak-link protocol would be applied
to both. Circulation and density profiles could then
be monitored over repeated realisations.

The spatial arrangement of the preparation requires
particular care. A fixed weak link can respond directly
to the position of a density or phase pattern, producing
a geometrical effect that might be mistaken for an
intrinsic phase-sensitive response. A nearly uniform
ring would avoid this complication; if a weak link
is used, the relative position of the prepared pattern
and the link should also be varied. The persistence
of the prepared relative phases must be checked over
the interval preceding the first phase slip.

The relevant observable would be the conditional
survival of a winding sector, rather than a single
deterministic lifetime. Let $\mathcal{P}$ label the
preparation and $\mathcal{F}$ the common future
protocol. If $T_n$ denotes the duration of one
continuous visit to winding sector $n$, the
corresponding survival function is
\begin{equation}
S_n(\tau\mid\mathcal{P},\mathcal{F})
=
\Pr\!\left(
T_n>\tau
\mid
\mathcal{P},\mathcal{F},
\text{entry into }n
\right),
\label{eq:sector_survival}
\end{equation}
where $\tau$ is the time elapsed since entry into
the sector. Separate visits following a back-slip
should be treated as distinct events, with the
direction of each exit recorded; observation-window
censoring should also be accounted for. A difference
between phase-conditioned survival functions would
provide a direct test of preparation-dependent
topological kinetics.

Several limitations remain. The holographic
calculations use a probe-limit Abelian--Higgs model,
a finite set of seeded modes and a restricted range
of backgrounds and driving protocols. We have not
reached a genuinely single-mode-dominant regime,
established a universal phase-dependent lifetime
law, or isolated a unique phase combination as
the cause of the response. The second background is
a retuned preparation and does not isolate charge
density as the only changed parameter. The
Gross--Pitaevskii calculation, in turn, is a
one-dimensional dissipative model and does not
provide a quantitative mapping to the holographic
time or to an experimental temperature. Finally,
the observed final winding numbers are finite-time
plateaux, not proofs of infinite-time attractors.

The point of this work is not that the equations
retain an independent dependence on the past once
the complete instantaneous state is specified.
Different histories prepare different complex
states, and information about those preparations
remains in the state on which the subsequent
nonlinear evolution acts. We have shown that this
information can remain visible in the timing of
individual topological transitions, even when the
future drive and the observed final winding are
the same. The route to a topological state has
its own dynamics, and its clock is not fixed by
the final winding alone.

\begin{acknowledgments}
The authors acknowledge HIAS for access to the ``Quantum Universe
Physical Simulation Platform''. B.-X.G. was supported by the
National Natural Science Foundation of China under Grant
No.~12505066, and Y.P. under Grant No.~12605094.
\end{acknowledgments}

\clearpage
\appendix

\section{Numerical Protocols, Convergence and Event Localisation}
\label{app:convergence}

We collect here the numerical checks supporting the nonlinear
results in the main text. They concern spatial and radial
resolution, time integration, and the localisation of winding
transitions. These tests address distinct sources of numerical
uncertainty and are treated separately.

\subsection{Construction of the matched history protocols}
\label{app:matched_protocols}

The three matched-history preparations in
Sec.~\ref{subsec:matched_histories} are designed to add the same
differential QNM action to the common base-rate ramp. We use
\(\Delta\gamma_{65}(v)\) as defined in
Sec.~\ref{subsec:qnm}, and set the reference action to
\begin{equation}
\mathcal W
=\Delta\gamma_{65}(v_h)\,\tau_{\rm ref},
\qquad
v_h=0.4086,
\qquad
\tau_{\rm ref}=20.
\end{equation}
The single-dwell protocol is the reference. The other two protocols
are constructed to add the same \(\mathcal W\), relative to the
base-rate ramp; the common ramps outside the writing window are
identical.

For the split-dwell protocol, define the two ends of the clean
writing window by
\begin{equation}
\begin{aligned}
\delta v_{\rm pad}
&=\max\!\left[5\times10^{-5},
0.06\bigl(v_{\rm crit}(-0.750)-v_{\rm crit}(-0.625)\bigr)\right],\\
v_1&=v_{\rm crit}(-0.625)+\delta v_{\rm pad},\\
v_2&=v_{\rm crit}(-0.750)-\delta v_{\rm pad}.
\end{aligned}
\end{equation}
Using the thresholds quoted in Sec.~\ref{subsec:history_protocol},
these endpoints are approximately \(v_1=0.40736\) and
\(v_2=0.40957\). The two dwell durations are chosen so that each contributes half of
the reference action:
\begin{equation}
\tau_1=\frac{\mathcal W/2}{\Delta\gamma_{65}(v_1)},
\qquad
\tau_2=\frac{\mathcal W/2}{\Delta\gamma_{65}(v_2)}.
\end{equation}
Both numerator and denominator are negative, so the dwell times are
positive. The system ramps at the base rate between the two holds.

For the continuous slow-window protocol, the velocity is swept from
\(v_1\) to \(v_2\) at a constant rate \(R_{\rm slow}<R\), with no
holding stage. Define
\begin{equation}
I=\int_{v_1}^{v_2}\!dv\,\Delta\gamma_{65}(v).
\end{equation}
Matching the additional differential action relative to the base
ramp gives
\begin{equation}
\mathcal W
=I\left(\frac{1}{R_{\rm slow}}-\frac{1}{R}\right),
\qquad
R_{\rm slow}
=\left(\frac{1}{R}+\frac{\mathcal W}{I}\right)^{-1},
\end{equation}
where \(R=4.25\times10^{-4}\) is the base ramp rate. Each protocol
then resumes the base-rate ramp to \(v_{\rm obs}=0.460\), followed
by the common three-unit quintic taper. The arrival imbalance and
residence times quoted in the main text are measured from the full
nonlinear evolution; they are not inferred from the QNM action.

For the second charge-density background, the velocity landmarks of
the original preparation are shifted by
\(\delta v=0.0134230186\). Thus
\begin{equation}
\begin{aligned}
v_i^{(2)}&=0.4084230186, &
 v_h^{(2)}&=0.4220230186,\\
v_{\rm obs}^{(2)}&=0.4734230186, &
 v_\star^{(2)}&=0.4383063373.
\end{aligned}
\end{equation}
The base ramp rate, dwell duration and taper duration are unchanged.
The near-degenerate crossing on this background was located from a
106-point scan of the QNM growth-rate difference and checked against
an independent continuation of the stationary branch.

For a transition from \(N=n\) to \(N=n-1\), we denote the
last reliable sample before the transition and the first
reliable sample after it by \(t_n^-\) and \(t_n^+\).
The event time is estimated by their midpoint,
\(t_{n\rightarrow n-1}=(t_n^-+t_n^+)/2\).
The interval \([t_n^-,t_n^+]\) is an event bracket, not a
statistical confidence interval. For a residence time
bounded by two transitions, a conservative localisation
bound is
\(\epsilon_T=(w_{\rm in}+w_{\rm out})/2\), where
\(w_{\rm in}\) and \(w_{\rm out}\) are the corresponding
bracket widths.

\subsection{Spatial resolution of the history-dependent trajectories}
\label{app:history_spatial_convergence}

We first revisit the three representative histories
\(\tau_h=0,20,40\), using
\(N_x=128,192,256\) at fixed \(N_z=28\).
The \(N_x=128\) and \(N_x=256\) results are the archived
production references, while \(N_x=192\) provides an
intermediate resolution.

For compact comparison, we define
\begin{equation}
\delta_{N_x,256}
=
\frac{T_{-4}^{N_x}-T_{-4}^{256}}
     {T_{-4}^{256}},
\qquad N_x=128,192.
\end{equation}
The residence times and relative differences are given
in Table~\ref{tab:history_spatial_convergence}.

\begin{table}[t]
    \centering
    \small
    \setlength{\tabcolsep}{3pt}
    \begin{tabular}{c|ccc|cc}
        \hline
        \(\tau_h\)
        & \(N_x=128\)
        & \(N_x=192\)
        & \(N_x=256\)
        & \(\delta_{128,256}\)
        & \(\delta_{192,256}\)
        \\
        \hline
        \(0\)
        & \(26.000000\)
        & \(26.000000\)
        & \(26.003018\)
        & \(-0.0116\%\)
        & \(-0.0116\%\)
        \\
        \(20\)
        & \(50.250000\)
        & \(50.600000\)
        & \(50.601180\)
        & \(-0.6940\%\)
        & \(-0.0023\%\)
        \\
        \(40\)
        & \(93.500000\)
        & \(94.900000\)
        & \(95.100014\)
        & \(-1.6825\%\)
        & \(-0.2103\%\)
        \\
        \hline
    \end{tabular}
    \caption{
    Spatial-resolution comparison for the three history
    preparations. All runs use \(N_z=28\). Each trajectory
    exhibits five resolved unit winding changes and reaches
    the same observed final sector, \(N_f=-5\). The relative
    differences are measured with respect to the
    \(N_x=256\) residence time.
    }
    \label{tab:history_spatial_convergence}
\end{table}

The ordering
\(T_{-4}(0)<T_{-4}(20)<T_{-4}(40)\) is unchanged
under refinement. The largest difference between
\(N_x=128\) and \(N_x=256\) is approximately \(1.68\%\),
for \(\tau_h=40\), while the \(N_x=192\) results are
already within \(0.21\%\) of the finest-resolution values.

The arrival modal imbalance is also insensitive to spatial
resolution. For each history, its variation across the
three resolutions is below \(4\times10^{-9}\).
The retained-edge Fourier-power diagnostic decreases
from approximately \(10^{-4}\) at \(N_x=128\), to
\(2\times10^{-6}\) at \(N_x=192\), and
\(5\times10^{-8}\) at \(N_x=256\).
The maximum constraint residual remains below
\(7.7\times10^{-12}\), and the relative charge drift
below \(5.1\times10^{-13}\) in the history-family runs.

The spatial refinement also changes the CFL-limited time
step. To examine time integration separately, we repeat
the \(\tau_h=20\), \(N_x=256\) calculation with the
Courant--Friedrichs--Lewy (CFL) factor reduced by one half.
The nominal step changes from
\begin{equation}
\Delta t_{\mathrm{CFL}}
=
8.78035\times10^{-3}
\quad\text{to}\quad
4.39018\times10^{-3}.
\end{equation}
The resulting residence times are
\begin{equation}
T_{-4}=50.6011798603
\qquad\text{and}\qquad
T_{-4}=50.6011798596,
\end{equation}
respectively. Their difference is approximately
\(6.6\times10^{-10}\).

The two transition times entering \(T_{-4}\) each shift
by approximately \(2.2\times10^{-3}\), whereas their
difference remains unchanged to the reported precision.
The final mean superflow is identical,
\(\bar u_{s,f}=0.3293707638\).

The refined run also passes the numerical-health checks,
with a maximum constraint residual below
\(7.7\times10^{-12}\). We do not assign a formal
convergence order to these event-derived times.
The spatial and time-step comparisons nevertheless show
that the principal history-dependent residence time is
stable under the refinements examined here.

\subsection{Radial resolution of the phase-only intervention}
\label{app:phase_radial_convergence}

We next examine the controlled phase-only intervention
discussed in Sec.~\ref{subsec:phase_intervention}. The complete
bulk QNM eigenvector of the \(m=-7\) mode is rotated by
\(\theta_{-7}(0)=\pm15^\circ\), while the initial boundary
amplitudes of all six seeded modes are kept fixed.

As a consistency check, the \(N_z=28\) runs agree with
the reference phase-only residence times to within
\(0.80\) simulation units, giving
\(T_{-4}=46.197832\) and \(56.001098\), respectively.

The initial-state construction is checked against the
reference initial-state routine through a zero-phase
comparison. The three tested initial fields, \(\Phi\),
\(A_x\) and \(\rho\), agree to floating-point precision.
The maximum relative error in the six measured boundary
seed amplitudes is below \(2.98\times10^{-11}\).

At each radial resolution, the stationary background and
all six QNM families are constructed independently.
The \(m=-7\) phase rotation is applied to the complete
complex bulk eigenvector before the real physical fields
are constructed. No bulk initial state is interpolated
between the two radial grids.

The resulting residence times are shown in
Table~\ref{tab:phase_radial_convergence}.

\begin{table}[t]
    \centering
    \small
    \setlength{\tabcolsep}{4pt}
    \begin{tabular}{ccccc}
        \hline
        \(N_z\)
        & \(T_{-4}(-15^\circ)\)
        & \(T_{-4}(+15^\circ)\)
        & \(\Delta T_{-4}\)
        & \(N_f\)
        \\
        \hline
        \(28\)
        & \(46.197832\)
        & \(56.001098\)
        & \(9.803265\)
        & \(-5\)
        \\
        \(32\)
        & \(46.197832\)
        & \(56.001098\)
        & \(9.803265\)
        & \(-5\)
        \\
        \hline
    \end{tabular}
    \caption{
    Radial-resolution comparison of the phase-only
    intervention at \(N_x=256\). Both preparations exhibit
    five monotonic unit winding changes and reach the same
    observed final sector. Here \(\Delta T_{-4}\) denotes
    \(T_{-4}(+15^\circ)-T_{-4}(-15^\circ)\).
    }
    \label{tab:phase_radial_convergence}
\end{table}

The two radial resolutions preserve the same winding
sequence and the same sign of the phase response.
The six QNM growth rates at the target velocity change
by less than \(1.8\times10^{-10}\) in relative terms
between \(N_z=28\) and \(N_z=32\). The two preparations remain closely matched in
\(\Delta_{56}^{\star}\) at both radial resolutions:
their difference is below \(10^{-8}\) for \(N_z=28\)
and below \(10^{-9}\) for \(N_z=32\).

The identical residence-time midpoints in
Table~\ref{tab:phase_radial_convergence} should not be
taken to imply exact equality of the underlying event
times. Both resolutions use an output interval of \(0.10\).
The corresponding residence-time localisation bounds
are approximately \(0.100974\) and \(0.105364\) for the
two phase preparations.

All four trajectories pass the numerical-health checks.
The maximum constraint residual is below
\(1.02\times10^{-11}\), the relative charge drift below
\(8.9\times10^{-16}\), and the boundary error below
\(3.2\times10^{-14}\).

This is a two-resolution radial stability check,
not a continuum extrapolation. It supports the persistence
of the phase response over the tested radial resolutions,
without assigning an exact radial error to the event times.

\subsection{Event localisation and time-integration refinement}
\label{app:phase_event_time_refinement}

We now separate the uncertainty associated with event
localisation from that associated with the PDE time
integrator. The two tests use the same reconstructed
phase-only implementation at \(N_x=256\) and \(N_z=32\).

First, the diagnostic output interval is reduced from
\(0.10\) to \(0.025\), with the PDE time-step prescription
and spectral-filter cadence unchanged. The resulting
residence times are compared in
Table~\ref{tab:phase_event_time_refinement}.

\begin{table}[t]
    \centering
    \scriptsize
    \setlength{\tabcolsep}{3pt}
    \begin{tabular}{lccccc}
        \hline
        Test
        & Output interval
        & \(\max\Delta t_{\mathrm{RK4}}\)
        & \(T_{-4}(-15^\circ)\)
        & \(T_{-4}(+15^\circ)\)
        & \(\Delta T_{-4}\)
        \\
        \hline
        Coarse event sampling
        & \(0.10\)
        & \(8.7804\times10^{-3}\)
        & \(46.197832\)
        & \(56.001098\)
        & \(9.803265\)
        \\
        Refined event sampling
        & \(0.025\)
        & \(8.7804\times10^{-3}\)
        & \(46.175881\)
        & \(56.001098\)
        & \(9.825216\)
        \\
        RK4 half-step
        & \(0.025\)
        & \(4.3902\times10^{-3}\)
        & \(46.175881\)
        & \(56.001098\)
        & \(9.825216\)
        \\
        \hline
    \end{tabular}
    \caption{
    Event-localisation and time-integration checks for the
    phase-only intervention at \(N_x=256\) and \(N_z=32\).
    The final column is the difference between the two
    phase preparations. The tabulated RK4 step is the
    largest actual step after the CFL restriction; the
    imposed step cap remains \(0.02\) in all three runs.
    }
    \label{tab:phase_event_time_refinement}
\end{table}

Reducing the output interval decreases the individual
residence-time localisation bounds from \(0.100974\)
and \(0.105364\) to \(0.026341\) for each phase
preparation. The corresponding bound on the phase
contrast is \(0.052682\).

The refined contrast is
\(\Delta T_{-4}=9.825216\), compared with
\(9.803265\) at the coarser output interval. The
difference, \(0.021951\), is smaller than the refined
contrast bracket bound. The positive phase response
is therefore unchanged by the improved event localisation.

We then test the time integrator independently.
At fixed \(N_x=256\), \(N_z=32\) and output interval
\(0.025\), each original RK4 step is replaced by two
consecutive half-steps. The outer time grid, protocol
event times and spectral-filter cadence in physical
time are preserved. The largest actual RK4 step
therefore decreases from
\(8.78035\times10^{-3}\) to
\(4.39018\times10^{-3}\).

The two phase-only residence times and their difference
are unchanged at the reported precision. Both trajectories
retain five monotonic unit winding changes and reach
\(N_f=-5\). The time-integration test also passes the
numerical-health and arrival-matching checks.

These tests address different sources of numerical
uncertainty: the first refines the event brackets, while
the second tests sensitivity to the PDE time integrator.
Neither is a continuum extrapolation.

\subsection{Spatial refinement of the common-future runs}
\label{app:common_future_validation}

We next summarise the spatial-refinement checks for the
common-future calculations in
Sec.~\ref{subsec:common_future}. The \(N_x=128\)
trajectories are compared with the \(N_x=256\)
production runs. Both resolutions use the same production
CFL prescription.

Table~\ref{tab:q3_spatial_refinement} gives the
penultimate-sector residence times whenever the winding
history permits an unambiguous comparison.

\begin{table}[t]
    \centering
    \scriptsize
    \setlength{\tabcolsep}{3pt}
    \begin{tabular}{cccccc}
        \hline
        \(v_c\)
        & \(\theta_{-7}(0)\)
        & \(N_f^{128}/N_f^{256}\)
        & Back-slip count \(128/256\)
        & \(T_{\mathrm{pen}}^{128}\)
        & \(T_{\mathrm{pen}}^{256}\)
        \\
        \hline
        \(0.420\)
        & \(-15^\circ\)
        & \(-5/-5\)
        & \(0/0\)
        & \(30.900\)
        & \(31.201\)
        \\
        \(0.420\)
        & \(+15^\circ\)
        & \(-5/-5\)
        & \(0/0\)
        & \(34.800\)
        & \(34.898\)
        \\
        \(0.430\)
        & \(-15^\circ\)
        & \(-6/-6\)
        & \(0/0\)
        & \(4.800\)
        & \(4.899\)
        \\
        \(0.430\)
        & \(+15^\circ\)
        & \(-6/-6\)
        & \(1/0\)
        & ---
        & \(1.901\)
        \\
        \(0.440\)
        & \(-15^\circ\)
        & \(-6/-6\)
        & \(1/0\)
        & ---
        & \(2.502\)
        \\
        \(0.440\)
        & \(+15^\circ\)
        & \(-6/-6\)
        & \(1/0\)
        & ---
        & \(1.901\)
        \\
        \hline
    \end{tabular}
    \caption{
    Spatial-refinement check for the common-future
    phase interventions. A dash indicates that the
    \(N_x=128\) trajectory contains a back-slip, so that
    no unambiguous penultimate-sector residence time
    is assigned. When \(N_f=-5\), the penultimate sector
    is \(N=-4\); when \(N_f=-6\), it is \(N=-5\).
    All \(N_x=256\) trajectories have monotonic unit
    winding histories.
    }
    \label{tab:q3_spatial_refinement}
\end{table}

At \(v_c=0.420\), both resolutions give the same
monotonic winding sequence, and the residence times
change by less than \(1\%\). At \(v_c=0.430\), the
\(-15^\circ\) trajectory also remains monotonic at
both resolutions. The \(+15^\circ\) trajectory,
however, contains one back-slip at \(N_x=128\);
this excursion is absent at \(N_x=256\).

The same distinction appears at \(v_c=0.440\).
Both lower-resolution trajectories contain one
back-slip, whereas both \(N_x=256\) trajectories
follow six monotonic unit winding changes and reach
\(N_f=-6\).

For completeness, the first entry into the
\(N=-4\) sector remains identifiable in these
lower-resolution runs, even when a later back-slip
prevents an unambiguous penultimate-sector assignment.
At \(v_c=0.430\), the \(+15^\circ\) first-entry time
changes from \(44.700\) to \(44.700782\). At
\(v_c=0.440\), the corresponding values change from
\(10.600\) to \(10.698861\) and from \(12.800\) to
\(12.898340\) for the \(-15^\circ\) and
\(+15^\circ\) preparations, respectively. These
first-entry times are descriptive and are not
substituted for \(T_{\mathrm{pen}}\).

The lower-resolution excursions are not reproduced
at \(N_x=256\) and are therefore not taken as evidence
for physical back-slips. All \(N_x=256\) cases satisfy
the numerical-health criteria monitored in these
calculations. This comparison supports the refined
winding histories, but does not establish a formal
spatial convergence order.

\subsection{Spatial refinement on the second background}
\label{app:second_density_validation}

Finally, we compare the phase-only trajectories on
the second charge-density background,
\(\bar{\rho}=10.543785\), at \(N_x=128\) and
\(N_x=256\), with \(N_z=28\) fixed. The lower-resolution
values are the archived Q4-B references; the higher
resolution is the subsequent refinement.

\begin{table}[t]
    \centering
    \small
    \setlength{\tabcolsep}{4pt}
    \begin{tabular}{ccccc}
        \hline
        \(\theta_{-7}(0)\)
        & \(T_{-4}^{128}\)
        & \(T_{-4}^{256}\)
        & Change
        & Back-slips \(128/256\)
        \\
        \hline
        \(-15^\circ\)
        & \(39.200\)
        & \(39.401838\)
        & \(0.515\%\)
        & \(0/0\)
        \\
        \(+15^\circ\)
        & \(46.800\)
        & \(47.602689\)
        & \(1.715\%\)
        & \(0/0\)
        \\
        \hline
    \end{tabular}
    \caption{
    Spatial-refinement comparison on the second
    charge-density background. Both resolutions give
    five reliable unit winding changes, the same
    observed final winding \(N_f=-5\), and one
    uninterrupted visit to \(N=-4\).
    }
    \label{tab:second_density_spatial_refinement}
\end{table}

The two resolutions preserve the same winding sequence
and the same sign of the phase response. The phase
contrast increases from \(7.600\) at \(N_x=128\)
to \(8.200851\) at \(N_x=256\). The event brackets
at the higher resolution give
\begin{equation}
7.998902
\leq
\Delta T_{-4}
\leq
8.402799,
\end{equation}
where
\(\Delta T_{-4}=T_{-4}(+15^\circ)-T_{-4}(-15^\circ)\).

The retained-edge Fourier-power fraction decreases
from \(1.60\times10^{-4}\) and \(1.27\times10^{-4}\)
at \(N_x=128\), to approximately
\(1.15\times10^{-7}\) for both preparations at
\(N_x=256\). The higher-resolution trajectories pass
the numerical-health checks and contain no detected
back-slips or non-unit winding changes.

The \(N_x=256\) runs end in a finite-time plateau.
We therefore treat \(N_f=-5\) as the observed final
winding, not as proof of an infinite-time attractor.

This is a two-resolution stability check, not a
formal continuum extrapolation. Moreover, the second
background is prepared with a shifted velocity protocol.
The comparison therefore does not isolate charge density
as the only changed physical parameter.

\section{Topological Diagnostics and Additional Nonlinear Checks}
\label{app:nonlinear_checks}

This appendix gives further details of the topological diagnostics
and nonlinear checks used in the main text. We describe the
reliability criteria and event bracketing, validate the winding
diagnostic on synthetic fields, and summarise the Fourier-tail
check, fixed-flow calibration, and local phase-slip maps.

\subsection{Boundary winding and measurement reliability}
\label{app:winding_reliability}

The boundary condensate is denoted by \({\cal O}(x,t)\) in the
main text. For compactness, we use \(\psi(x,t)\equiv{\cal O}(x,t)\)
when discussing the numerical diagnostics and the local field maps.
Numerically, it is extracted from the boundary radial derivative
of the bulk scalar field, consistently with the boundary
quantisation specified in Sec.~\ref{sec:setup}.

The winding number itself is defined in
Sec.~\ref{sec:setup}. In addition to its integer value, the
diagnostic retains the unrounded quantity
\begin{equation}
N_{\rm raw}(t)
=
\frac{1}{2\pi}\sum_{n=0}^{N_x-1}\delta\theta_n(t),
\end{equation}
and its distance from the nearest integer,
\begin{equation}
r_N(t)=|N_{\rm raw}(t)-N(t)|.
\end{equation}

We also monitor the minimum condensate amplitude relative to
its spatial median,
\begin{equation}
D_{\min}(t)
=
\frac{\min_n|\psi_n(t)|}
{\operatorname{median}_n|\psi_n(t)|}.
\end{equation}
The numerical phase measurement is regarded as reliable only
when both
\begin{equation}
D_{\min}>10^{-8},
\qquad
r_N\leq5\times10^{-6}
\end{equation}
are satisfied. The first condition excludes samples at which
the condensate is too strongly depleted for its phase to be
reliably determined, while the second checks the consistency
of the link-phase sum with an integer winding.

These are numerical reliability criteria, not physical
thresholds for the occurrence of a phase slip. In particular,
a small \(D_{\min}\) does not by itself establish a topological
transition.

\subsection{Event bracketing and condensate depletion}
\label{app:event_detection}

A candidate topological event is identified by comparing
successive reliable winding measurements. If their integer
values differ, the transition is bracketed by the last reliable
sample before the change and the first reliable sample after it.
Intermediate samples at which the phase is unreliable are
excluded from the winding comparison, but their condensate
depletion values are retained.

For each bracket, the diagnostic records the minimum value of
\(D_{\min}\) encountered between its two endpoints. This gives
independent evidence of local condensate depletion during the
transition. The event detector itself does not impose a fixed
depletion threshold: the topological criterion is the change
between two reliable integer windings.

The fixed-flow calibration includes an additional numerical
acceptance check. At least one bracket must contain a sample
with
\begin{equation}
D_{\min}\leq0.25.
\end{equation}
This is a run-management criterion used to ensure that the
calibration resolves a sufficiently deep depletion. It is
not a universal definition of a phase slip and is not used
to assign the winding number.

A jump with \(|\Delta N|>1\) requires particular care.
It may represent a genuinely multiple-winding transition,
or several transitions occurring within one output interval.
The diagnostic does not automatically split such a jump
into separate unit events. A higher output cadence is needed
before making that distinction.

The event midpoint and the uncertainty associated with its
bracket are defined in Appendix~\ref{app:convergence}. Thus,
the integer winding establishes the topological change,
while the local amplitude and event bracket describe what
can be resolved about its occurrence.

\subsection{Validation of the diagnostic implementation}
\label{app:diagnostic_tests}

Before its use in the nonlinear evolution, the winding
diagnostic was tested independently on synthetic periodic
fields. The test grid has \(N_x=128\) points and
\(L_x=16\pi\). The complete test suite contains 14 checks,
all of which pass.

The tests and their scope are summarised in
Table~\ref{tab:winding_diagnostic_tests}.

\begin{table}[t]
    \centering
    \small
    \setlength{\tabcolsep}{4pt}
    \begin{tabular}{lcc}
        \hline
        Test & Number & Result \\
        \hline
        Exact integer windings
        & 7 & Pass \\
        Global phase invariance
        & 1 & Pass \\
        Periodic translation invariance
        & 1 & Pass \\
        Principal-phase branch cut
        & 1 & Pass \\
        Weak-noise robustness
        & 1 & Pass \\
        Mean-superflow convention
        & 1 & Pass \\
        Synthetic event bracketing
        & 1 & Pass \\
        Fourier spectral-tail check
        & 1 & Pass \\
        \hline
        Total
        & 14 & 14/14 \\
        \hline
    \end{tabular}
    \caption{
    Unit tests of the phase and winding diagnostic.
    The exact-winding tests use \(N=0,\pm1,\pm2,\pm5\).
    The weak-noise test comprises 100 independent
    perturbations of a field with \(N=-5\).
    }
    \label{tab:winding_diagnostic_tests}
\end{table}

The global-phase test checks invariance under four independent
constant phase rotations, while the translation test uses five
periodic shifts of the spatial grid. The branch-cut test uses
a field with \(N=5\) whose principal phase repeatedly crosses
the endpoints of the interval \((-\pi,\pi]\). The measured
winding remains unchanged.

For the 100 weak-noise realisations, the largest distance
from an integer is \(8.88\times10^{-16}\), and the minimum
depletion ratio remains above \(0.9615\). Thus, the test
does not rely on configurations approaching a vanishing
condensate.

The mean-superflow check uses \(\mu=6\), \(v=0.415\) and
\(L_x=16\pi\). It verifies the homogeneous relation
\(\bar u_s=-a_x/\mu=v\) for \(N=0\), and the corresponding
winding contribution for \(N=2\). Both are reproduced
to better than \(10^{-14}\).

Finally, a synthetic event is constructed with an
intentionally unreliable intermediate sample, for which
\(D_{\min}=2\times10^{-4}\). The detector correctly
brackets one transition with \(\Delta N=-1\), using
the reliable winding measurements on either side.
The spectral-tail test, described below, separately
checks the response to low- and high-frequency Fourier modes.

These tests establish the internal consistency of the
diagnostic routines. They do not replace the spatial
and temporal refinement tests of the nonlinear PDE
evolution presented in Appendix~\ref{app:convergence}.

\subsection{Fourier spectral-tail diagnostic}
\label{app:spectral_tail}

A Fourier spectral filter is employed during the nonlinear
evolution, retaining modes within
\begin{equation}
|m|\leq m_{\rm cut},
\qquad
m_{\rm cut}=\lfloor N_x/3\rfloor.
\end{equation}
To monitor the spectral content near the upper edge of
the retained band, we define the Fourier coefficients
\begin{equation}
\widehat{\psi}_m(t)
=
\frac{1}{N_x}
\sum_{n=0}^{N_x-1}
\psi_n(t)e^{-2\pi i mn/N_x}.
\end{equation}

The corresponding retained-edge power fraction is
\begin{equation}
P_{\rm edge}(t)
=
\frac{
\displaystyle
\sum_{0.8m_{\rm cut}<|m|\leq m_{\rm cut}}
|\widehat{\psi}_m(t)|^2
}{
\displaystyle
\sum_m|\widehat{\psi}_m(t)|^2
}.
\end{equation}
This quantity measures the fraction of Fourier power
concentrated near the upper edge of the retained band.
It is used to monitor the spectral resolution of the
nonlinear evolution and to compare results obtained
at different spatial resolutions. No universal
pass/fail threshold is imposed.

The corresponding convergence results are reported in
Appendix~\ref{app:convergence}.

\subsection{Fixed-flow phase-slip calibration}
\label{app:fixed_flow_calibration}

We next summarise the independent fixed-flow calibration
used to test the nonlinear winding diagnostic. The
background has
\begin{equation}
\bar\rho=10.0417,
\qquad
v=0.415,
\qquad
L_x=16\pi.
\end{equation}
A single Fourier mode with
$
k=-0.5
$
is seeded with amplitude
$
\epsilon=10^{-5}.
$
The corresponding unstable QNM is used to construct the
initial perturbation, including its scalar and gauge-field
components.

The evolution is performed at \(N_z=28\), with
\(N_x=128\) and \(256\). The nominal RK4 step is bounded
by \(\Delta t_{\max}=0.02\) and the Fourier CFL condition.
The diagnostic output interval is \(0.25\). The
two-thirds dealiasing filter is applied every four
RK4 steps. The early-time growth fit uses the interval
\(20\leq t\leq280\), before the evolution becomes
strongly nonlinear.

The unstable QNM growth rate is
\begin{equation}
\operatorname{Im}\omega_{\rm QNM}
=
0.00526610528,
\end{equation}
whereas the early-time fit to the nonlinear evolution gives
\begin{equation}
\gamma_{\rm fit}=0.00526617.
\end{equation}
The relative difference is approximately
\(1.23\times10^{-5}\). Thus, the initial evolution
reproduces the linear QNM growth before entering the
strongly nonlinear regime.

At both spatial resolutions, the evolution exhibits
the same net winding change,
\begin{equation}
N:0\longrightarrow-4,
\end{equation}
together with strong local condensate depletion. The
final mean gauge-invariant flow is
\begin{equation}
\bar u_{s,f}=0.3381473915
\end{equation}
at both resolutions.

The numerical comparison is summarised in
Table~\ref{tab:fixed_flow_calibration}.

\begin{table}[t]
    \centering
    \small
    \setlength{\tabcolsep}{4pt}
    \begin{tabular}{ccc}
        \hline
        Quantity
        & \(N_x=128\)
        & \(N_x=256\)
        \\
        \hline
        Final winding
        & \(-4\)
        & \(-4\)
        \\
        Final mean flow
        & \(0.3381473915\)
        & \(0.3381473915\)
        \\
        \(P_{\rm edge}^{\max}\)
        & \(1.05\times10^{-4}\)
        & \(4.48\times10^{-8}\)
        \\
        Maximum constraint residual
        & \(6.55\times10^{-12}\)
        & \(6.70\times10^{-12}\)
        \\
        \hline
    \end{tabular}
    \caption{
    Fixed-flow nonlinear calibration at \(v=0.415\).
    Both resolutions give the same net winding change
    and final mean flow, while the retained-edge Fourier
    power decreases substantially at higher resolution.
    }
    \label{tab:fixed_flow_calibration}
\end{table}

The transition \(0\rightarrow-4\) is quoted here as a
net winding change. The available diagnostic does not
resolve it into four temporally separate unit phase-slip
events, and no individual transition times are assigned.
The purpose of this calculation is to establish the
connection between linear QNM growth, nonlinear
condensate depletion and a subsequent change of winding,
rather than to provide a separate history-dependent
relaxation law.

\subsection{Local phase and amplitude structure}
\label{app:local_slip_structure}

The local maps shown in Fig.~\ref{fig:local_phase_slip}
give a closer view of the phase-slip region. They display
the boundary condensate \(\psi\equiv{\cal O}\) for three
initial phases of the \(m=-7\) mode,
\begin{equation}
\theta_{-7}(0)=-15^\circ,\qquad
0^\circ,\qquad
+15^\circ.
\end{equation}

The upper row shows \(\arg\psi(x,t)\), while the lower
row shows \(|\psi(x,t)|\). The coordinates are centred
on the event, using \(x-x_{\rm slip}\) and
\(t-t_{\rm slip}\), with the white cross marking
\(x=x_{\rm slip}\) and \(t=t_{\rm slip}\).

The phase maps display a local rearrangement of the
phase texture across the event region. The corresponding
amplitude maps show a narrow channel of reduced
condensate amplitude in the same spacetime region.
This correspondence is visible in all three preparations,
although the surrounding phase pattern and local details
differ. The three amplitude panels use a common colour
scale, so that their depletion profiles can be compared
directly.

The phase colour scale is periodic: \(-\pi\) and \(+\pi\)
represent the same phase. Their identical endpoint colours
are therefore intentional and do not indicate a physical
discontinuity. The maps are constructed from the
\(N_x=8192\) local replay.

These plots illustrate the local field structure, but
they do not replace the winding diagnostic. In particular,
a low-amplitude channel does not by itself prove that
\(|\psi|\) vanishes exactly at a point, nor does the
colour pattern establish a spacetime topological charge.
The topological classification used in the holographic
evolution remains the change in the independently
measured boundary winding. We do not assign a plaquette
charge to these holographic field maps.

\section{Maxwell Feedback and Multimode Source Decomposition}
\label{app:maxwell_feedback}

This appendix gives the bulk equations and Fourier-space
details underlying the Maxwell--scalar feedback discussed
in Sec.~VI. We first write the evolution equations in the
variables used by the numerical solver, then decompose the
\(q=+\kappa\) component of the longitudinal Maxwell equation.
Finally, we relate the selected boundary mode combination
to its bulk counterpart. The decomposition identifies an
allowed phase-sensitive coupling channel, but does not
single out a dominant source term.

\subsection{Bulk equations and field conventions}
\label{app:maxwell_equations}

The bulk scalar \(\Psi\) was introduced in
Sec.~II. For the time-dependent evolution, we use the
rescaled field
\begin{equation}
\varphi(t,z,x)=\frac{\Psi(t,z,x)}{z}.
\end{equation}
With the source-free boundary condition
\(\Psi=z^2\Psi_++\cdots\), this gives
\begin{equation}
\varphi(t,0,x)=0,
\qquad
{\cal O}(x,t)=\left.\partial_z\varphi\right|_{z=0},
\label{eq:appC-boundary-condensate}
\end{equation}
where \({\cal O}\) is the boundary condensate defined
in Sec.~II. Thus \(\varphi\) is the bulk field evolved
by the numerical solver, whereas \({\cal O}\) is its
boundary counterpart.

In ingoing Eddington--Finkelstein coordinates and radial
gauge, \(A_z=0\), the scalar equation becomes
\begin{equation}
\begin{aligned}
\partial_t\partial_z\varphi
={}&\frac12\partial_z(f\partial_z\varphi)
+\frac12\partial_x^2\varphi
-iA_x\partial_x\varphi+iA_t\partial_z\varphi\\
&-\frac{i}{2}(\partial_xA_x-\partial_zA_t)\varphi
-\frac12(z+A_x^2)\varphi .
\end{aligned}
\label{eq:appC-scalar}
\end{equation}
Here \(f(z)\) is the blackening factor defined in
Sec.~II.

The longitudinal Maxwell equation is
\begin{equation}
\partial_t\partial_z A_x
=
\frac12\partial_z(f\partial_z A_x)
-|{\varphi}|^2A_x
+\operatorname{Im}(\varphi^*\partial_x\varphi)
+\frac12\partial_x\partial_z A_t .
\label{eq:appC-maxwell}
\end{equation}
The temporal gauge field is reconstructed from the radial
Maxwell constraint,
\begin{equation}
\partial_z^2 A_t
=
\partial_z\partial_x A_x
-2\operatorname{Im}(\varphi^*\partial_z\varphi).
\label{eq:appC-constraint}
\end{equation}

At the boundary, the conditions are
\begin{equation}
\begin{gathered}
\varphi(t,0,x)=0,\qquad
A_t(t,0,x)=\mu(t),\qquad
A_x(t,0,x)=a_x(t),\\
\partial_z A_t(t,0,x)=-\rho(t,x).
\end{gathered}
\end{equation}
Both boundary gauge sources are homogeneous in \(x\).
The constraint for \(A_t\) is solved at every
Runge--Kutta stage, rather than evolving \(A_t\) as
an independent field.

It is useful to introduce the spatial gauge-covariant
derivative
\begin{equation}
D_x=\partial_x-iA_x.
\end{equation}
The two scalar terms in Eq.~\eqref{eq:appC-maxwell}
involving \(\partial_x\varphi\) and \(A_x\) can then
be combined as
\begin{equation}
-|{\varphi}|^2A_x
+\operatorname{Im}(\varphi^*\partial_x\varphi)
=
\operatorname{Im}(\varphi^*D_x\varphi).
\end{equation}
For the source decomposition below, however, these
contributions are kept separate, as they are in the
numerical evaluation.

\subsection{Fourier-resolved Maxwell source}
\label{app:maxwell_fourier}

The periodic domain has length \(L_x=16\pi\), with
fundamental wavenumber
\begin{equation}
\kappa=\frac{2\pi}{L_x}=\frac18,
\end{equation}
as in Sec.~VI B. For a bulk field \(F(z,x,t)\), its
Fourier coefficient is
\begin{equation}
F_q(z,t)
=
\frac{1}{L_x}\int_0^{L_x}dx\,
F(z,x,t)e^{-iqx}.
\label{eq:appC-fourier}
\end{equation}
We also write \([F]_q\equiv F_q\).

We focus on the \(q=+\kappa\) harmonic. The right-hand
side of Eq.~\eqref{eq:appC-maxwell} is separated into
four contributions:
\begin{equation}
\partial_t\partial_z A_{x,\kappa}
=
{\cal M}_{x,\kappa}
=
{\cal S}_{{\rm rad},\kappa}
+{\cal S}_{{\rm cur},\kappa}
+{\cal S}_{{\rm gd},\kappa}
+{\cal S}_{{\rm el},\kappa}.
\label{eq:appC-source-decomposition}
\end{equation}
Here
\begin{equation}
\begin{aligned}
{\cal S}_{{\rm rad},\kappa}
&=\frac12\partial_z(f\partial_z A_{x,\kappa}),\\
{\cal S}_{{\rm cur},\kappa}
&=[\operatorname{Im}(\varphi^*\partial_x\varphi)]_\kappa,\\
{\cal S}_{{\rm gd},\kappa}
&=-[|\varphi|^2A_x]_\kappa,\\
{\cal S}_{{\rm el},\kappa}
&=\frac{i\kappa}{2}\partial_z A_{t,\kappa}.
\end{aligned}
\label{eq:appC-source-terms}
\end{equation}
The labels denote the radial, bare-current,
gauge-density and electric contributions, respectively.
The last three terms are coupled through the scalar
field and the Maxwell constraint; they are not
independent physical sources.

To make the mode coupling explicit, expand the bulk
scalar as
\begin{equation}
\varphi(z,x,t)
=
\sum_{m\in\mathbb Z}\varphi_m(z,t)e^{ik_mx},
\qquad
k_m=m\kappa,
\label{eq:appC-bulk-fourier}
\end{equation}
where \(\varphi_m(z,t)\) denotes the radial Fourier
profile of the rescaled scalar. The density harmonic
is then
\begin{equation}
[|\varphi|^2]_\kappa
=
\sum_m\varphi_m\varphi_{m-1}^*,
\label{eq:appC-density-harmonic}
\end{equation}
and the bare-current contribution becomes
\begin{equation}
{\cal S}_{{\rm cur},\kappa}
=
\frac12\sum_m(k_m+k_{m-1})
\varphi_m\varphi_{m-1}^*.
\label{eq:appC-current-harmonic}
\end{equation}

The gauge-density term involves an additional
convolution with the gauge field:
\begin{equation}
[|\varphi|^2A_x]_\kappa
=
\sum_p[|\varphi|^2]_{\kappa-p}A_{x,p},
\label{eq:appC-gauge-density}
\end{equation}
where \(p\) runs over the allowed Fourier wavenumbers
\(p=n\kappa\), with \(n\in\mathbb Z\). It therefore
depends on the full set of populated Fourier sectors,
rather than on the density harmonic at \(q=\kappa\)
alone.

The radial constraint at this harmonic reads
\begin{equation}
\partial_z^2 A_{t,\kappa}
=
i\kappa\partial_z A_{x,\kappa}
-2[\operatorname{Im}(\varphi^*\partial_z\varphi)]_\kappa.
\label{eq:appC-fourier-constraint}
\end{equation}
Since the boundary chemical potential is homogeneous,
the corresponding boundary conditions are
\begin{equation}
A_{t,\kappa}(0,t)=0,
\qquad
\partial_z A_{t,\kappa}(0,t)=-\rho_\kappa(t),
\label{eq:appC-fourier-boundary}
\end{equation}
where
\begin{equation}
\rho_\kappa(t)
=
\frac{1}{L_x}\int_0^{L_x}dx\,
\rho(t,x)e^{-i\kappa x}
\end{equation}
is the \(+\kappa\) Fourier coefficient of the
boundary charge density. Thus the electric contribution
is determined by the evolving scalar and gauge fields;
it is not an independently imposed source.

The four terms in Eq.~\eqref{eq:appC-source-decomposition}
are complex radial profiles. They must be added as
complex quantities before the norm of the total source
is evaluated. Adding their individual magnitudes would
discard the cancellations between them.

\subsection{The selected three-mode contribution}
\label{app:triad_coupling}

The selected boundary combination \(B_\kappa^{(567)}\)
and the phase \(\Phi_{567}\) were introduced in
Sec.~VI B. They involve the equally spaced modes
\(m=-5,-6,-7\) and display the phase dependence of
their contribution to the \(+\kappa\) harmonic of
the boundary condensate density.

The corresponding bulk contribution is constructed
from the radial Fourier profiles:
\begin{equation}
\widetilde B_\kappa^{(567)}(z,t)
=
\varphi_{-5}\varphi_{-6}^*
+\varphi_{-6}\varphi_{-7}^*.
\label{eq:appC-bulk-triad}
\end{equation}
This bulk quantity should not be identified with
the boundary combination. In particular, the
boundary Fourier coefficients satisfy
\begin{equation}
{\cal O}_{k_m}(t)
=
\left.\partial_z\varphi_m(z,t)\right|_{z=0}.
\label{eq:appC-boundary-mode}
\end{equation}
The boundary contribution is therefore built from
radial derivatives at the AdS boundary, not from
the values of the bulk profiles at an arbitrary
radial position.

Neither \(B_\kappa^{(567)}\) nor
\(\widetilde B_\kappa^{(567)}\) is the complete
density harmonic. The latter contains all neighbouring
Fourier pairs, as shown in
Eq.~\eqref{eq:appC-density-harmonic}, while the
corresponding boundary harmonic is
\begin{equation}
[|{\cal O}|^2]_\kappa
=
\sum_m{\cal O}_{k_m}{\cal O}_{k_{m-1}}^*.
\label{eq:appC-boundary-density}
\end{equation}
Contributions from the remaining modes can therefore
reinforce or oppose the selected three-mode term.
A change in the boundary combination need not be
reflected in the same way in the complete boundary
density harmonic, nor does it determine the full
bulk Maxwell source.

\subsection{Maxwell--scalar feedback}
\label{app:maxwell_feedback_loop}

The same neighbouring-mode products appear in the
bare-current term, with their respective wavenumber
weights, and in the gauge-density term, through its
convolution with \(A_x\). The resulting response of
\(A_x\), together with the constraint-determined
\(A_t\), then feeds back into the scalar equation
through its gauge-covariant derivative and
field-strength terms.

The feedback is nonlinear and involves the full
Fourier spectrum and its radial structure. The
selected three-mode combination provides a useful
way of displaying one phase-sensitive contribution,
but the scalar evolution does not close on these
three modes. In particular, Eq.~\eqref{eq:appC-scalar}
contains no independent local
\(|\varphi|^2\varphi\) self-interaction term. The
nonlinear coupling considered here instead arises
through the gauge fields and their response to the
scalar sources.

The phase-only preparation also does not vary
\(\Phi_{567}\) in isolation. Rotating the complete
initial bulk QNM eigenvector of the \(m=-7\) mode
changes several relative phase combinations involving
that mode. The resulting evolution therefore tests
sensitivity to the complex modal preparation, rather
than the response to a single independently controlled
triad phase.

\subsection{Scope of the decomposition}
\label{app:maxwell_budget_interpretation}

The four terms in Eq.~\eqref{eq:appC-source-decomposition}
are coupled parts of a single Maxwell equation, not
independent physical sources or separate positive
contributions to an energy budget. In particular,
the electric term remains tied to the scalar and gauge
fields through the radial constraint. Nor is
\(B_\kappa^{(567)}\) the complete density harmonic or
the full bulk Maxwell source: the corresponding
quantity in Eq.~\eqref{eq:appC-bulk-triad} retains
only two of the neighbouring-mode products, while
the bare-current and gauge-density terms carry,
respectively, their wavenumber weights and the
additional convolution in
Eq.~\eqref{eq:appC-gauge-density}.

The phase dependence of this selected combination
therefore identifies one channel through which the
initial modal phases can enter the gauge-field
response, but does not establish its dominance.
Nor does the decomposition identify a common
net-source threshold or a critical value of
\(\Phi_{567}\) preceding the final phase slip.
We therefore do not attribute the residence-time
difference to any unique Maxwell source term.

\section{An Independent Dissipative Gross--Pitaevskii Model}
\label{app:dgpe_bridge}

We give further details of the independent one-dimensional
superfluid model used to test whether phase-sensitive
topological relaxation also occurs outside the holographic
setting. The model is not a reduction of the holographic
equations. It provides a separate setting in which the
complex amplitudes of unstable modes can be prepared
directly, without reconstructing an earlier driving history.

We focus on the initial-state construction, the independent
phase coordinates and the numerical checks on the resulting
topological trajectories. The model has a uniform ring,
a fixed drift and phenomenological dissipation, with no
weak link or spatially varying potential.

\subsection{Model and physical conventions}
\label{app:dgpe_model}

We consider a complex order parameter
\(\psi_{\rm d}(x,t)\) on a periodic ring of length
\(L_{\rm d}=16\pi\). Its evolution is governed by
\begin{equation}
(i-\Gamma_{\rm d})\partial_t\psi_{\rm d}
=
\left[
-\frac12\partial_x^2
-i u_{\rm d}\partial_x
+g_{\rm d}(|\psi_{\rm d}|^2-1)
\right]\psi_{\rm d}.
\label{eq:dgpe}
\end{equation}
Here \(i\) is the imaginary unit, \(i^2=-1\);
\(\Gamma_{\rm d}\) is a phenomenological damping
coefficient, \(g_{\rm d}\) is the contact interaction
strength, and \(u_{\rm d}\) is a constant drift parameter
in the rotating frame. We use
\begin{equation}
\Gamma_{\rm d}=0.20,\qquad
g_{\rm d}=1,\qquad
u_{\rm d}=1.35.
\end{equation}

The homogeneous field \(\psi_{\rm d}=1\) is a stationary
solution. Dissipation makes the energetic instability
of this state dynamically accessible through growing
Bogoliubov modes. The mean particle density is not
conserved by this evolution, which is effectively
grand-canonical.

The corresponding rotating-frame grand-potential
functional is
\begin{equation}
{\cal G}_{\rm d}[\psi_{\rm d}]
=
\int_0^{L_{\rm d}}dx\,
\left[
\frac12|\partial_x\psi_{\rm d}|^2
+u_{\rm d}\operatorname{Re}
\bigl(\psi_{\rm d}^*(-i\partial_x)\psi_{\rm d}\bigr)
+\frac{g_{\rm d}}{2}(|\psi_{\rm d}|^2-1)^2
\right].
\label{eq:dgpe_grand_potential}
\end{equation}

We define the mean density, momentum per unit length
and grand-potential density by
\begin{equation}
\bar n_{\rm d}
=
\frac{1}{L_{\rm d}}\int_0^{L_{\rm d}}dx\,|\psi_{\rm d}|^2,
\qquad
p_{\rm d}
=
\frac{1}{L_{\rm d}}\int_0^{L_{\rm d}}dx\,
\operatorname{Re}\bigl[\psi_{\rm d}^*(-i\partial_x)\psi_{\rm d}\bigr],
\qquad
\bar{\cal G}_{\rm d}
=
\frac{{\cal G}_{\rm d}}{L_{\rm d}}.
\end{equation}

For the deterministic equation,
\begin{equation}
\frac{d{\cal G}_{\rm d}}{dt}
=
-\frac{2\Gamma_{\rm d}}{1+\Gamma_{\rm d}^2}
\int_0^{L_{\rm d}}dx\,
\left|
\frac{\delta{\cal G}_{\rm d}}{\delta\psi_{\rm d}^*}
\right|^2
\leq0.
\end{equation}
Thus \({\cal G}_{\rm d}\) is a Lyapunov functional
of the deterministic evolution.

The drift \(u_{\rm d}\) and the time variable in
Eq.~\eqref{eq:dgpe} belong to this independent model.
They are not numerically comparable to the holographic
superfluid velocity or its time units. Nor is the
present equation calibrated to a particular cold-atom
experiment.

\subsection{Bogoliubov modes and initial preparation}
\label{app:dgpe_bdg}

We expand around the homogeneous state using six
negative Fourier modes,
\begin{equation}
{\cal M}=\{-3,-4,-5,-6,-7,-8\},
\qquad
q_m=\frac{2\pi m}{L_{\rm d}}=\frac{m}{8},
\qquad
h_m=\frac{q_m^2}{2}.
\end{equation}
Here \(m\) is the integer Fourier-mode label. The
linearised equation for the particle--hole pair
\((\delta\psi_{q_m},\delta\psi_{-q_m}^*)\) is determined
by
\begin{equation}
{\mathsf L}_m=
\begin{pmatrix}
\displaystyle\frac{h_m+u_{\rm d}q_m+g_{\rm d}}{i-\Gamma_{\rm d}}
&
\displaystyle\frac{g_{\rm d}}{i-\Gamma_{\rm d}}
\\[6pt]
\displaystyle\frac{g_{\rm d}}{-i-\Gamma_{\rm d}}
&
\displaystyle\frac{h_m-u_{\rm d}q_m+g_{\rm d}}{-i-\Gamma_{\rm d}}
\end{pmatrix}.
\label{eq:dgpe_bdg_matrix}
\end{equation}

For each mode, we select the eigenvalue \(\lambda_m\)
with the largest real part and normalise its eigenvector
to \((1,\beta_m)^T\):
\begin{equation}
{\mathsf L}_m
\begin{pmatrix}1\\\beta_m\end{pmatrix}
=
\lambda_m
\begin{pmatrix}1\\\beta_m\end{pmatrix}.
\end{equation}
The corresponding growth rates are listed in
Table~\ref{tab:dgpe_bdg_growth}.

\begin{table}[t]
    \centering
    \small
    \setlength{\tabcolsep}{7pt}
    \begin{tabular}{ccc}
        \hline
        \(m\) & \(q_m\) & \(\operatorname{Re}\lambda_m\) \\
        \hline
        \(-3\) & \(-0.375\) & \(0.049887915\) \\
        \(-4\) & \(-0.500\) & \(0.055960023\) \\
        \(-5\) & \(-0.625\) & \(0.059067460\) \\
        \(-6\) & \(-0.750\) & \(0.060166613\) \\
        \(-7\) & \(-0.875\) & \(0.059632652\) \\
        \(-8\) & \(-1.000\) & \(0.057559022\) \\
        \hline
    \end{tabular}
    \caption{
    Linear growth rates of the six seeded BdG modes
    at \(\Gamma_{\rm d}=0.20\), \(g_{\rm d}=1\) and
    \(u_{\rm d}=1.35\). The central modes form a
    closely competing group.
    }
    \label{tab:dgpe_bdg_growth}
\end{table}

The initial field is constructed from the complete
particle--hole eigenvectors:
\begin{equation}
\psi_{\rm d}(x,0)
=
1+\sum_{m\in{\cal M}}
\left[
c_m e^{iq_mx}
+(c_m\beta_m)^*e^{-iq_mx}
\right],
\qquad
c_m=a_{\rm seed}e^{i\theta_m},
\qquad
a_{\rm seed}=0.018.
\label{eq:dgpe_initial_state}
\end{equation}
Here \(\theta_m\) is the phase of the negative-\(m\)
particle coefficient, while the positive-\(m\) hole
component is fixed by the corresponding BdG eigenvector.

All six one-sided particle amplitudes have the same
magnitude. Their individual Fourier powers are
\begin{equation}
|c_m|^2=a_{\rm seed}^2=3.24\times10^{-4}.
\end{equation}
The hole components are included in the initial state
and are rotated consistently with their particle
partners. A change of modal phase therefore leaves
both the particle and hole Fourier powers unchanged.

The linearisation was checked directly against the
numerical field evolution. For a perturbation of
amplitude \(10^{-7}\), the relative error in the
linearised field response is \(7.32\times10^{-7}\).
The test was performed at \(N_x=128\) and \(256\),
where \(N_x\) denotes the number of spatial Fourier
grid points.

\subsection{Independent phase coordinates}
\label{app:dgpe_phase_coordinates}

The six non-zero particle coefficients have phases
\begin{equation}
\boldsymbol{\theta}
=
(\theta_{-3},\theta_{-4},\theta_{-5},
\theta_{-6},\theta_{-7},\theta_{-8})^T.
\end{equation}
The homogeneous component has phase \(\theta_0\),
fixed to zero in the chosen initial gauge. All phase
coordinates are understood modulo \(2\pi\).

Four independent second phase differences are defined by
\begin{equation}
\Phi_j=\theta_{j-1}-2\theta_j+\theta_{j+1},
\qquad j=-4,-5,-6,-7.
\end{equation}
For the central three-mode group, we write
\begin{equation}
\Phi_{567}\equiv\Phi_{-6}
=\theta_{-5}-2\theta_{-6}+\theta_{-7}.
\end{equation}

A fifth coordinate retains the phase information
relative to the homogeneous background:
\begin{equation}
\Xi=2\theta_{-3}-\theta_{-6}-\theta_0.
\end{equation}
The five coordinates are collected in
\begin{equation}
\mathbf Q=
(\Phi_{-4},\Phi_{-5},\Phi_{567},\Phi_{-7},\Xi)^T.
\end{equation}

Before fixing the global phase, these combinations
are invariant under a global \(U(1)\) rotation of the
complete field and under a rigid spatial translation.
The inclusion of \(\theta_0\) in \(\Xi\) is important:
rotating only the six non-zero modal coefficients while
holding the homogeneous background fixed is not a
global phase rotation of the complete initial state.
It changes the relative phase between the seeded modes
and the background.

There are seven phases when the homogeneous component
is included. Removing the global phase and spatial
translation leaves five independent phase coordinates.
In the gauge \(\theta_0=0\), their relation to the six
modal phases is
\begin{equation}
\mathbf Q={\mathsf C}_5\boldsymbol{\theta},
\qquad
{\mathsf C}_5=
\begin{pmatrix}
1&-2&1&0&0&0\\
0&1&-2&1&0&0\\
0&0&1&-2&1&0\\
0&0&0&1&-2&1\\
2&0&0&-1&0&0
\end{pmatrix}.
\label{eq:dgpe_phase_matrix}
\end{equation}

The matrix \({\mathsf C}_5\) has rank five. Its
Moore--Penrose pseudoinverse, denoted by
\({\mathsf C}_5^+\), satisfies
\begin{equation}
{\mathsf C}_5{\mathsf C}_5^+={\mathsf I}_5,
\end{equation}
where \({\mathsf I}_5\) is the \(5\times5\) identity
matrix. The null direction in the six-dimensional
modal-phase space is proportional to
\((-3,-4,-5,-6,-7,-8)^T\), corresponding to a rigid
translation of the ring. The pseudoinverse fixes the
representative orthogonal to this direction.

The reference preparation is
\begin{equation}
\boldsymbol{\theta}_{\rm ref}
=(0,0,0,0,-15^\circ,0)^T,
\qquad
\mathbf Q_{\rm ref}
=(0,0,-15^\circ,30^\circ,0)^T.
\end{equation}
We write \(\delta\mathbf Q=\mathbf Q-\mathbf Q_{\rm ref}\)
for a phase kick relative to this reference state.
It is implemented through
\begin{equation}
\delta\boldsymbol{\theta}
={\mathsf C}_5^+\delta\mathbf Q,
\qquad
\boldsymbol{\theta}
=\boldsymbol{\theta}_{\rm ref}
+\delta\boldsymbol{\theta}.
\end{equation}
The numerical residual in the right-inverse identity
is below \(10^{-13}\).

This construction identifies independent phase
directions within the selected finite set of initial
BdG amplitudes. It does not imply that the subsequent
nonlinear field evolution is described by five variables.

\subsection{Equal-grand-potential phase intervention}
\label{app:dgpe_equal_energy}

We compare two preparations, A and B, that differ
only in the \(\Phi_{567}\) coordinate. All kicks are
measured relative to the reference state defined above.
Preparation A receives a \(+45^\circ\) kick, while
preparation B receives a \(-20.593347^\circ\) kick.
The latter is found by bisection on the initial
grand potential, before either nonlinear trajectory
is evolved. The other four phase coordinates are
held fixed.

We denote the integer winding by \(N_{\rm d}\).
For a single uninterrupted visit to sector \(s\),
\(T_s^{\rm d}\) denotes the time between entry into
and exit from that sector; the event-time convention
is specified below.

At \(N_x=512\), the two preparations have identical
particle and hole Fourier powers, initial mean density,
momentum and winding. In particular,
\begin{equation}
\bar n_{\rm d}^{A}(0)=\bar n_{\rm d}^{B}(0)
=1.002471300442754,
\qquad
p_{\rm d}^{A}(0)=p_{\rm d}^{B}(0)
=-1.0168292163752115\times10^{-3}.
\end{equation}
Both have initial winding \(N_{\rm d}=0\). Their
initial grand-potential densities differ by only
\(2.17\times10^{-19}\).

The principal comparison is summarised in
Table~\ref{tab:dgpe_equal_energy_results}.

\begin{table}[t]
    \centering
    \small
    \setlength{\tabcolsep}{5pt}
    \begin{tabular}{cccc}
        \hline
        A kick
        & B kick
        & Grand-potential mismatch
        & \(T_{-5}^{A}/T_{-5}^{B}\)
        \\
        \hline
        \(+45^\circ\)
        & \(-20.593347^\circ\)
        & \(2.17\times10^{-19}\)
        & \(4.685866/14.585583\)
        \\
        \hline
    \end{tabular}
    \caption{
    Equal-initial-grand-potential intervention along
    \(\Phi_{567}\), at \(N_x=512\). The kicks are
    measured relative to the common reference state.
    }
    \label{tab:dgpe_equal_energy_results}
\end{table}

Both trajectories follow the same monotonic sequence
of unit winding changes,
\begin{equation}
N_{\rm d}:0\rightarrow-1\rightarrow-2
\rightarrow-3\rightarrow-4\rightarrow-5\rightarrow-6,
\end{equation}
and reach the same observed final sector,
\(N_{\rm d}=-6\). Their continuous residence times
in \(N_{\rm d}=-5\) differ by a factor of
\begin{equation}
\frac{T_{-5}^{B}}{T_{-5}^{A}}=3.1127.
\end{equation}

The comparison is deliberately restricted to one
phase direction. It shows that matching the initial
modal powers, mean density, momentum and grand
potential does not fix the subsequent residence time
in a winding sector.

\subsection{Numerical diagnostics and resolution checks}
\label{app:dgpe_numerics}

The winding is evaluated using the link-phase sum,
\begin{equation}
N_{\rm d}(t)=
\operatorname{round}\left[
\frac{1}{2\pi}
\sum_{n=0}^{N_x-1}
\operatorname{Arg}
\left(\psi_{{\rm d},n}^*\psi_{{\rm d},n+1}\right)
\right],
\end{equation}
where \(x_n=nL_{\rm d}/N_x\),
\(\psi_{{\rm d},n}=\psi_{\rm d}(x_n,t)\), and the
spatial index is periodic. Here \(\operatorname{Arg}\)
denotes the principal complex argument.

The winding is checked after every integration step.
A transition is bracketed by the two adjacent time
steps at which the integer winding differs. Its
recorded event time is the midpoint of that bracket.
For a single uninterrupted visit to sector \(s\),
\(T_s^{\rm d}\) is the difference between the midpoints
of the entering and leaving events. Separate visits
following a back-slip are never combined.

As a local check, we also evaluate the phase winding
around a space--time plaquette. For vertices ordered as
\((t_j,x_n)\rightarrow(t_j,x_{n+1})
\rightarrow(t_{j+1},x_{n+1})
\rightarrow(t_{j+1},x_n)\), its signed charge is
\begin{equation}
Q_\square=
\operatorname{round}\left[
\frac{1}{2\pi}
\sum_{(a,b)\in\partial\square}
\operatorname{Arg}(\psi_{{\rm d},a}^*\psi_{{\rm d},b})
\right].
\end{equation}
With this orientation,
\begin{equation}
Q_\square=-\Delta N_{\rm d},
\qquad
\Delta N_{\rm d}
=N_{\rm d}^{\rm after}-N_{\rm d}^{\rm before}.
\end{equation}
Bilinear interpolation of the four complex field
values on an event plaquette also provides a subgrid
estimate of an internal zero of the interpolated field.
This is an interpolation check, not an independent
continuum solution for the location of an exact zero.

Spatial derivatives are evaluated using a Fourier
pseudospectral method, and the evolution is integrated
with fourth-order Runge--Kutta. The cubic nonlinearity
is projected onto the retained band
\begin{equation}
|m|\leq m_{\rm ret},
\qquad
m_{\rm ret}=\lfloor N_x/4\rfloor-2.
\end{equation}
This conservative one-half-rule truncation suppresses
cubic aliasing into the retained modes.

The retained-edge Fourier-power fraction is monitored
through
\begin{equation}
P_{\rm edge}^{\rm d}(t)
=
\frac{
\displaystyle\sum_{0.8m_{\rm ret}\leq|m|\leq m_{\rm ret}}
|\widehat{\psi}_m(t)|^2
}{
\displaystyle\sum_{|m|\leq m_{\rm ret}}
|\widehat{\psi}_m(t)|^2
},
\qquad
\widehat{\psi}_m(t)
=
\frac{1}{L_{\rm d}}
\int_0^{L_{\rm d}}dx\,
\psi_{\rm d}(x,t)e^{-iq_mx}.
\end{equation}

The resolution comparison for the principal
equal-energy pair is given in
Table~\ref{tab:dgpe_spatial_convergence}.

\begin{table}[t]
    \centering
    \small
    \setlength{\tabcolsep}{6pt}
    \begin{tabular}{ccc}
        \hline
        \(N_x\)
        & \(T_{-5}^{A}\)
        & \(T_{-5}^{B}\)
        \\
        \hline
        \(128\) & \(4.664667\) & \(14.773945\) \\
        \(256\) & \(4.630000\) & \(14.610000\) \\
        \(512\) & \(4.685866\) & \(14.585583\) \\
        \hline
    \end{tabular}
    \caption{
    Resolution comparison for the principal
    equal-energy \(\Phi_{567}\) pair. The production
    time-step prescription also changes with \(N_x\).
    }
    \label{tab:dgpe_spatial_convergence}
\end{table}

At \(N_x=256\), halving the nominal time step from
\(0.010\) to \(0.005\) leaves the two residence times
unchanged at the reported precision: \(4.630\) and
\(14.610\), respectively. The event sequence is also
unchanged.

At \(N_x=512\), the actual time step is
\(\Delta t=0.0059998286\). The maximum retained-edge
power fractions in the two principal phase controls
are below \(1.5\times10^{-13}\). The sampled grand
potential does not increase along either deterministic
trajectory.

Selected trajectories were extended to \(t=300\).
The principal pair remains in the observed sector
\(N_{\rm d}=-6\) for approximately \(230.973\) and
\(220.005\) time units after its last winding change,
for A and B, respectively. These are long finite-time
plateaux, not proofs of infinite-time attractors.

\subsection{Scope of the cross-model comparison}
\label{app:dgpe_scope}

The dGPE calculation provides an independent example
of phase-sensitive winding-sector kinetics in a model
with dynamics distinct from the holographic
Maxwell--scalar system. Changing the phase structure
of the initial BdG preparation changes the residence
time in an intermediate winding sector, even when
the initial modal powers, mean density, momentum and
grand potential are matched.

The five phase coordinates identify independent
directions within the selected finite-mode preparation.
They are not a complete set of variables for the
subsequent nonlinear evolution. The present comparison
tests one such direction, \(\Phi_{567}\), and does not
establish that it is the only phase coordinate relevant
to the dynamics.

The two models have different microscopic dynamics.
The dGPE contains contact interactions and
phenomenological damping, whereas the holographic
evolution involves bulk Maxwell--scalar feedback.
The present calculation establishes neither a
one-to-one correspondence between these mechanisms
nor a quantitative prediction for a particular
cold-atom experiment. Its role is more limited:
phase-sensitive topological relaxation can also occur
in a distinct, explicitly specified dissipative
superfluid model.

\bibliography{Ref}

@article{Landau1941,
    author = "Landau, L. D.",
    title = "{Theory of the Superfluidity of Helium II}",
    journal = "Phys. Rev.",
    volume = "60",
    number = "4",
    pages = "356--358",
    year = "1941",
    doi = "10.1103/PhysRev.60.356"
}

@article{Herzog2009,
    author = "Herzog, Christopher P. and Kovtun, Pavel K. and Son, Dam T.",
    title = "{Holographic model of superfluidity}",
    eprint = "0809.4870",
    archivePrefix = "arXiv",
    primaryClass = "hep-th",
    journal = "Phys. Rev. D",
    volume = "79",
    number = "6",
    pages = "066002",
    year = "2009",
    doi = "10.1103/PhysRevD.79.066002"
}

@article{Arean2010,
    author = "Are{\'a}n, Daniel and Basu, Pallab and Krishnan, Chethan",
    title = "{The Many Phases of Holographic Superfluids}",
    eprint = "1006.5165",
    archivePrefix = "arXiv",
    primaryClass = "hep-th",
    journal = "JHEP",
    volume = "10",
    pages = "006",
    year = "2010",
    doi = "10.1007/JHEP10(2010)006"
}

@article{Amado2009,
    author = "Amado, Irene and Kaminski, Matthias and Landsteiner, Karl",
    title = "{Hydrodynamics of Holographic Superconductors}",
    eprint = "0903.2209",
    archivePrefix = "arXiv",
    primaryClass = "hep-th",
    journal = "JHEP",
    volume = "05",
    pages = "021",
    year = "2009",
    doi = "10.1088/1126-6708/2009/05/021"
}

@article{Amado2014,
    author = "Amado, Irene and Are{\'a}n, Daniel and Jim{\'e}nez-Alba, Amadeo and Landsteiner, Karl and Melgar, Luis and Salazar Landea, Ignacio",
    title = "{Holographic Superfluids and the Landau Criterion}",
    eprint = "1307.8100",
    archivePrefix = "arXiv",
    primaryClass = "hep-th",
    journal = "JHEP",
    volume = "02",
    pages = "063",
    year = "2014",
    doi = "10.1007/JHEP02(2014)063"
}

@article{Gouteraux2023,
    author = "Gout{\'e}raux, Blaise and Sottovia, Filippo and Mefford, Eric",
    title = "{Critical superflows and thermodynamic instabilities in superfluids}",
    eprint = "2212.10410",
    archivePrefix = "arXiv",
    primaryClass = "hep-th",
    journal = "Phys. Rev. D",
    volume = "108",
    number = "8",
    pages = "L081903",
    year = "2023",
    doi = "10.1103/PhysRevD.108.L081903"
}

@article{Arean2024,
    author = "Are{\'a}n, Daniel and Gout{\'e}raux, Blaise and Mefford, Eric and Sottovia, Filippo",
    title = "{Hydrodynamics and instabilities of relativistic superfluids at finite superflow}",
    eprint = "2312.08243",
    archivePrefix = "arXiv",
    primaryClass = "hep-th",
    journal = "JHEP",
    volume = "05",
    pages = "272",
    year = "2024",
    doi = "10.1007/JHEP05(2024)272"
}

@article{Keranen2010,
    author = "Ker{\"a}nen, Ville and Keski-Vakkuri, Esko and Nowling, Sean and Yogendran, K. P.",
    title = "{Inhomogeneous Structures in Holographic Superfluids: I. Dark Solitons}",
    eprint = "0911.1866",
    archivePrefix = "arXiv",
    primaryClass = "hep-th",
    journal = "Phys. Rev. D",
    volume = "81",
    number = "12",
    pages = "126011",
    year = "2010",
    doi = "10.1103/PhysRevD.81.126011"
}

@article{Yang2021,
    author = "Yang, Peng and Li, Xin and Tian, Yu",
    title = "{Instability of holographic superfluids in optical lattice}",
    eprint = "2109.09080",
    archivePrefix = "arXiv",
    primaryClass = "hep-th",
    journal = "JHEP",
    volume = "11",
    pages = "190",
    year = "2021",
    doi = "10.1007/JHEP11(2021)190"
}

@article{Lan2025,
    author = "Lan, Shanquan and Liu, Hong and Tian, Yu and Zhang, Hongbao",
    title = "{Landau instability and soliton formations}",
    eprint = "2010.06232",
    archivePrefix = "arXiv",
    primaryClass = "hep-th",
    journal = "Phys. Rev. D",
    volume = "112",
    number = "2",
    pages = "L021901",
    year = "2025",
    doi = "10.1103/g8sg-prdf"
}

@article{Varoquaux2015,
    author = "Varoquaux, Eric",
    title = "{Anderson's considerations on the flow of superfluid helium: Some offshoots}",
    eprint = "1406.5629",
    archivePrefix = "arXiv",
    primaryClass = "cond-mat.other",
    journal = "Rev. Mod. Phys.",
    volume = "87",
    number = "3",
    pages = "803--854",
    year = "2015",
    doi = "10.1103/RevModPhys.87.803"
}

@article{LiShiZhang2022,
    author = "Li, Zhi-Hong and Shi, Han-Qing and Zhang, Hai-Qing",
    title = "{Holographic topological defects in a ring: role of diverse boundary conditions}",
    eprint = "2111.15230",
    archivePrefix = "arXiv",
    primaryClass = "hep-th",
    journal = "JHEP",
    volume = "05",
    pages = "056",
    year = "2022",
    doi = "10.1007/JHEP05(2022)056"
}

@article{Yan2026,
    author = "Yan, Chang-Xu and Chang, Hao-Ran and Gao, Meng and Tian, Yu and Zhang, Hongbao",
    title = "{Formation of holographic vortex in a rotating shell-shaped superfluid}",
    eprint = "2607.07172",
    archivePrefix = "arXiv",
    primaryClass = "hep-th",
    journal = "JHEP",
    volume = "09",
    pages = "198",
    year = "2026",
    doi = "10.1007/JHEP09(2026)198"
}

@article{Ge:2026nyg,
    author = "Ge, Bo-Xuan",
    title = "{History-Dependent Mode Selection in a Driven Holographic Superfluid}",
    eprint = "2608.14997",
    archivePrefix = "arXiv",
    primaryClass = "hep-th",
    year = "2026"
}

\end{document}